\documentclass[a4paper,fleqn,usenatbib,useAMS]{mpazh}
\pdfoutput=1

\usepackage{hyperref}

\usepackage[utf8]{inputenc}
\usepackage{graphicx}
\usepackage{grffile}
\usepackage{natbib}

\usepackage{floatrow}
\usepackage{caption}
\usepackage{longtable}

\usepackage{multirow}

\renewcommand{\theenumi}{\roman{enumi}}
\def\doubleline{\vskip 3pt\hrule \vskip 1.5pt \hrule \vskip 5pt}

\def\ps1{\emph{Pan-STARRS1}}

\def\srg{\textit{SRG}}
\def\art{ART-XC}
\def\ero{eROSITA}
\def\rosat{\textit{ROSAT}}
\def\xmm{\textit{XMM-Newton}}
\def\swift{\textit{Swift}}
\def\integral{\textit{INTEGRAL}}

\def\nh{N_{\rm H}}
\def\lx{L_{\rm X}}

\def\tub{T$\ddot{\rm U}$BITAK}

\def\azt{AZT-33IK}
\def\rtt{RTT-150}

\title{New Active Galactic Nuclei Detected by the \art\
and \ero\ Telescopes Onboard the \srg\ Observatory during an All-Sky
X-ray Survey}
\author[Uskov et. al.]{
	G.S. Uskov,
			\thanks{\href{mailto:uskov@cosmos.ru}{\nolinkurl{uskov@cosmos.ru}}}
				$^{1}$
		I. A. Zaznobin,
				$^{1}$
		S. Yu. Sazonov,
				$^{1}$
		A. N. Semena,
				$^{1}$
		M. R. Gilfanov,
				$^{1,2}$
		\newauthor
		R. A. Burenin,
				$^{1}$
			M. V. Eselevich,
				$^{3}$
		R. A. Krivonos,
				$^{1}$
		A. R. Lyapin,
				$^{1}$
		P. S. Medvedev,
				$^{1}$
		\newauthor
		G. A. Khorunzhev,
				$^{1}$
		R. A. Sunyaev,
				$^{1,2}$
	\\
			$^{1}$Space Research Institute, Russian Academy of Sciences, Moscow,
117997 Russia\\
			$^{2}$Max Planck Institut for Astrophysik, Karl-Schwarzschild-Str. 1,
85741 Garching, Germany\\
			$^{3}$Institute of Solar--Terrestrial Physics, Russian Academy of
Sciences, Siberian Branch, Irkutsk, 664033 Russia}

\date{Accepted XXX. Received YYY; in original form ZZZ}

\pubyear{2022}

\begin{document}

\label{firstpage}
\pagerange{\pageref{firstpage}--\pageref{lastpage}}
\maketitle

\begin{abstract}
We present the results of our identification of 17 X-ray sources
detected in the 4--12 keV energy range by the Mikhail Pavlinsky \art\
telescope during the first year of the \srg\ all-sky survey. Three of
them have been discovered by the \art\ telescopes, while the remaining
ones have already been known previously as X-ray sources, but their
nature has remained unknown. We took optical spectra for nine sources
located in the northern sky (\(\delta> -20^\circ\)) with the 1.6-m
AZT-33IK telescope at the Sayan Observatory (the Institute of
Solar--Terrestrial Physics, the Siberian Branch of the Russian Academy
of Sciences) and the 1.5-m Russian--Turkish telescope at the
\tub~National Observatory. For the remaining objects we have analyzed
the archival optical spectra taken during the 6dF survey. All of the
investigated objects have turned out to be Seyfert galaxies (eight of
type 1, seven of type 2, and two of intermediate type 1.8) at redshifts
up to \(z \approx 0.15\). Based on data from the \ero\ and \art\
telescopes onboard the \srg\ observatory, we have obtained X-ray spectra
in the energy range 0.2--20 keV for eight sources. A significant
intrinsic absorption (\(\nh > 10^{22}\) cm\(^{-2}\)) has been detected
in three of them, with two of them being probably strongly absorbed
(\(\nh \sim 10^{23}\) cm\(^{-2}\)). This paper is a continuation of the
series of publications on the optical identification of active galactic
nuclei detected by the \art~telescope.
\end{abstract}


\section{Introduction}
\label{s:intro}

The Spectrum--Roentgen--Gamma (\srg) orbital observatory \citep{sunyaev2021} has conducted an all-sky X-ray survey since December 12, 2019.
A total of eight surveys, each with a duration of six months, are planned to be conducted. There are two telescopes with grazing-incidence X-ray optics onboard the \srg~observatory: \ero~\citep{predehl2021} and
Mikhail Pavlinsky \art~\citep{pavlinsky2021} operating in the 0.2--8
and 4--30 keV energy bands, respectively.

The second \srg~sky survey was completed on December 15, 2020. The first
catalog of sources detected in the 4--12 keV energy band by the
\art~telescope was produced from the sum of the two surveys \citep{pavlinsky2022}. This catalog was correlated with: (1) the catalogs of
sources detected in previous X-ray sky surveys, (2) the preliminary
catalog of sources detected on the half of the celestial sphere
\(0<|l|<180^\circ\)\footnote{Russian scientists are responsible for processing the 
\ero\ data in this part of the sky.} in the soft X-ray energy band
during the first three \srg/\ero~all-sky surveys, and (3) the catalogs
of astrophysical objects in other wavelength ranges. As a result, a list
of objects discovered by the \art~telescope and previously known X-ray
sources confirmed with the \art~telescope, whose nature remained unknown
or poorly studied, was compiled. Most of these objects were also
detected by the \ero~telescope onboard the \srg~observatory.
Spectroscopic observations are carried out at ground-based optical
telescopes to establish the nature of these objects.

In this paper we present the results of our identification of nine
sources from the first \art~catalog of sources using observations with
the 1.6-m \azt~telescope at the Sayan Observatory (the Institute of
Solar--Terrestrial Physics, the Siberian Branch of the Russian Academy
of Sciences) and the 1.5-m Russian--Turkish telescope (\rtt) at the
\tub~National Observatory. These objects, located in the northern sky,
have turned out to be active galactic nuclei (AGNs). We also present the
results of our identification of eight \art~sources in the southern sky
based on the available archival data from the spectroscopic 6dF survey,
which have also turned out to be AGNs. In addition, we constructed
broadband (0.2--20\textasciitilde keV) X-ray spectra for eight northern
objects (with coordinates in the range \(0<|l|<180^\circ\)) from the
\ero~and \art~data during the first three half-year sky surveys. Using
these spectra, we have managed to reveal a significant intrinsic
absorption in several objects. This paper continues the series of
publications on the identification of new AGNs and cataclysmic variables
from the \srg/\art~allsky survey begun in \cite{zaznobin2021, zaznobin2022}
The presented luminosity estimates are based on the model of a flat
Universe with parameters \(H_0=70\) and \(\Omega_m = 0.3\).

\begin{table*}
    \caption{
    The X-ray sources for which the observations at the $\azt$ and $\rtt$ telescopes were carried out
    }
    \label{tab:list_src}
    \vskip 2mm
    \renewcommand{\arraystretch}{1.1}
    \renewcommand{\tabcolsep}{0.25cm}
    \centering
    \footnotesize
    \begin{tabular}[t]{rccccccccl}
        \noalign{\doubleline}
    &  &\multispan2\hfil \ero\ coordinates \hfil & \multispan2\hfil Optical coordinates \hfil &  \\
 No. & \art\ source              & $\alpha$    & $\delta$     & $\alpha$    & $\delta$    &$\rm r_{\rm A}$&$\rm r_{\rm e}$& $F_{\rm A}^{4-12}$ & Discovered by\\
  \noalign{\vskip 3pt\hrule\vskip 5pt}
 1 & SRGA\,J$025234.3\!+\!431004$ & $-$         & $-$           & $43.14170$  & $+43.16740$  & $3.5\arcsec$  & $-$           & $2.7_{-1.8}^{+2.3}$ & \swift\\
 2 & SRGA\,J$062627.2\!+\!072734$ & $-$         & $-$           & $96.61250$  & $+7.45806$   & $5.8\arcsec$  & $-$           & $1.7_{-2.3}^{+3.6}$  & \rosat\\
 3 & SRGA\,J$070636.4\!+\!635109$ & $106.64528$ & $+63.84891$   & $106.64500$ & $+63.84889$  & $16.8\arcsec$ & $1.4\arcsec$  & $5.4_{-2.6}^{+3.6}$ & \srg\\
 4 & SRGA\,J$092021.6\!+\!860249$ & $140.06928$ & $+86.05057$   & $140.06973$ & $+86.05012$  & $12.5\arcsec$ & $2.3\arcsec$  & $5.1_{-2.0}^{+2.5}$  & \rosat\\
 5 & SRGA\,J$195702.4\!+\!615036$ & $299.25991$ & $+61.84267$   & $299.26000$ & $+61.84306$  & $1.0\arcsec$  & $0.4\arcsec$  & $3.4_{-1.2}^{+1.4}$  & \rosat\\
 6 & SRGA\,J$221913.2\!+\!362014$ & $334.81076$ & $+36.33471$   & $334.81050$ & $+36.33630$  & $16.3\arcsec$ & $4.4\arcsec$  & $5.4_{-2.3}^{+3.0}$  & \srg\\
 7 & SRGA\,J$223714.9\!+\!402939$ & $339.31426$ & $+40.49534$   & $339.31458$ & $+40.49583$  & $9.1\arcsec$  & $1.5\arcsec$  & $5.0_{-2.3}^{+2.9}$    & \rosat\\
 8 & SRGA\,J$232037.8\!+\!482329$ & $350.16453$ & $+48.39126$   & $350.16417$ & $+48.39056$  & $16.2\arcsec$ & $1.0\arcsec$  & $1.6_{-1.5}^{+2.0}$ & \rosat\\
 9 & SRGA\,J$235250.6\!-\!170449$ & $358.21433$ & $-17.07735$   & $358.21417$ & $-17.07694$  & $16.6\arcsec$ & $1.4\arcsec$  & $6.6_{-3.0}^{+4.0}$  & \swift\\
\noalign{\vskip 3pt\hrule\vskip 5pt}
\end{tabular}
  \begin{flushleft}
  Column1: the ordinal sourcenumber inthe sample beingstudied. Column2: the source name in the \art\ catalog (the coordinates
of the X-ray sources used in the names are given for epoch J2000.0). Columns 3 and 4: the coordinates of the source from the
\ero\ data. Columns 5 and 6: the coordinates of the putative optical counterpart. Column 7: the distance between the positions
of the \art\ source and the optical counterpart. Column 8: the distance between the positions of the \ero\ source and the
optical counterpart. Column 9: the X-ray flux in the 4–12 keV energy band based on data from the first two \art\ sky surveys \citep{pavlinsky2022}, in units of $10^{-12}$~erg~s$^{-1}$~cm$^{-2}$. Column 10: the X-ray observatory that detected the source for the first time.
The \ero\ coordinates are not given for SRGA\,J$025234.3\!+\!431004$ due to its insufficiently significant detection by the \ero\
telescope and for SRGA\,J$062627.2\!+\!072734$ due to its location on the half of the sky $180<|l|<360^\circ$, on which German scientists
are responsible for processing the \ero\ data.
  \end{flushleft}
\end{table*}

\begin{table*}
    \caption{The X-ray sources for which archival 6dF data are available}
    \label{tab:list_src_6df}
    \vskip 2mm
    \renewcommand{\arraystretch}{1.1}
    \renewcommand{\tabcolsep}{0.35cm}
    \centering
    \footnotesize
    \begin{tabular}[t]{rlccccl}
        \noalign{\doubleline}
     &  &\multispan2\hfil Optical coordinates \hfil &  \\
 No. & \art\ source            & $\alpha$    & $\delta$    &$\rm r_{\rm A}$& $F_{\rm A}^{4-12}$   & Discovered by \\
  \noalign{\vskip 3pt\hrule\vskip 5pt}
 10  & SRGA\,J$030838.1\!-\!552041$ & $47.15875$  & $-55.34472$  & $4.0\arcsec$  & $4.8_{-1.5}^{+1.9}$  & \srg\\
 11  & SRGA\,J$052959.8\!-\!340157$ & $82.49669$  & $-34.03293$  & $7.5\arcsec$  & $4.6_{-1.7}^{+2.0}$  & \xmm\\
 12  & SRGA\,J$055053.7\!-\!621457$ & $87.72339$  & $-62.24863$  & $2.2\arcsec$  & $1.4_{-0.5}^{+0.5}$  & \rosat\\
 13  & SRGA\,J$060241.1\!-\!595152$ & $90.67472$  & $-59.86456$  & $6.3\arcsec$  & $2.5_{-0.8}^{+0.9}$  & \xmm\\
 14  & SRGA\,J$061322.9\!-\!290027$ & $93.35120$  & $-29.00633$  & $19.0\arcsec$ & $12.4_{-3.2}^{+3.9}$ & \rosat\\
 15  & SRGA\,J$063324.9\!-\!561424$ & $98.36091$  & $-56.23914$  & $14.8\arcsec$ &  $3.4_{-1.2}^{+1.4}$ & \rosat\\
 16  & SRGA\,J$064421.5\!-\!662620$ & $101.09111$ & $-66.43886$  & $2.2\arcsec$  & $0.6_{-0.5}^{+0.5}$  & \rosat\\
 17  & SRGA\,J$072823.5\!-\!440823$ & $112.09742$ & $-44.14005$  & $1.6\arcsec$  & $4.6_{-1.9}^{+2.5}$  & \rosat\\
\noalign{\vskip 3pt\hrule\vskip 5pt}
\end{tabular}
  \begin{flushleft}
  The contents of the columns are analogous to those in Table \ref{tab:list_src}. The sources are located on the half of the sky $180<|l|<360^\circ$ and, therefore, no information on the \ero\ data is given. 
  \end{flushleft}
\end{table*}

\section*{THE SAMPLE OF OBJECTS}

The sample of 17 objects being studied was selected from the catalog of X-ray sources detected by the \art\ telescope during the first year of its all- sky survey (December 12, 2019–December 15, 2020) \citep{pavlinsky2022}. We considered only point sources from this catalog. According to the catalog production criterion, all such sources were detected at a significance level of no less than 4.82 standard deviations in the 4–12 keV energy band. Based on the \art\ data, we measured the positions of the sources in the sky with an accuracy better than 30 arcsec. For eight sources from this sample located in the sky region $0<|l|<180^\circ$ there are also \srg/\ero\ data at our disposal, which allowed broadband X-ray spectra for these objects to be constructed from the set of \ero\ and \art\ data. The \ero\ data for the remaining nine sources belong to the German \srg/\ero\ consortium and are not considered here.

For nine of the 17 X-ray sources located in the northern sky ($\delta>-20^\circ$) we took optical spectra. For the remaining eight sources located in the southern sky we analyzed the available archival data from the spectroscopic 6dF survey of galaxies \citep{jones2004}. For all objects Tables~\ref{tab:list_src} and \ref{tab:list_src_6df} give: the coordinates of the source from the \art\ and \ero\ data (if available), the coordinates of the putative optical counterpart, the distance between the positions of the X-ray source and the optical counterpart, and the X-ray flux in the 4--12~keV energy band.
 
\section{X-RAY OBSERVATIONS}

At present, all of the sources from the sample have been observed during the first three \srg\ all-sky surveys. Using the combined \ero\ and \art\ data from these surveys, we constructed the spectra of eight sample sources located in the sky region $0<|l|<180^\circ$ in the energy range from 0.2 to 20~keV. The variability of the objects was not investigated.

The \art\ X-ray spectra were obtained with the ARTPRODUCTS v0.9 software \citep{pavlinsky2021} using the calibration files of version 20200401. The data from all seven \art\ modules were combined. The spectra of the sources were extracted in a region of radius 120\arcsec\ in three energy bands: 4--8, 8--12, and 12--20 keV.

The \ero\ data were processed with the calibration and data processing system created and maintained at the Space Research Institute of the Russian Academy of Sciences that was constructed using the elements of the eSASS (\ero\ Science Analysis Software System) package and the software developed by the science group on the X-ray catalog of the Russian \ero\ consortium. We extracted the source spectra in a circle of radius $R=60\arcsec$ and the background spectra in a ring with the inner radius $R_{\rm in}=120\arcsec$ and the outer radius $R_{\rm out}=300\arcsec$ around the source. If other sources fell into the background region, then they were excluded with the radius $R=40\arcsec$. The spectra were extracted from the data of all seven \art\ modules in the energy range 0.2–-9.0~keV. When fitting the spectra, the data were binned in such a way that there were at least three counts in each energy channel.

\begin{figure}
  \centering
    \includegraphics[width=0.85\columnwidth]{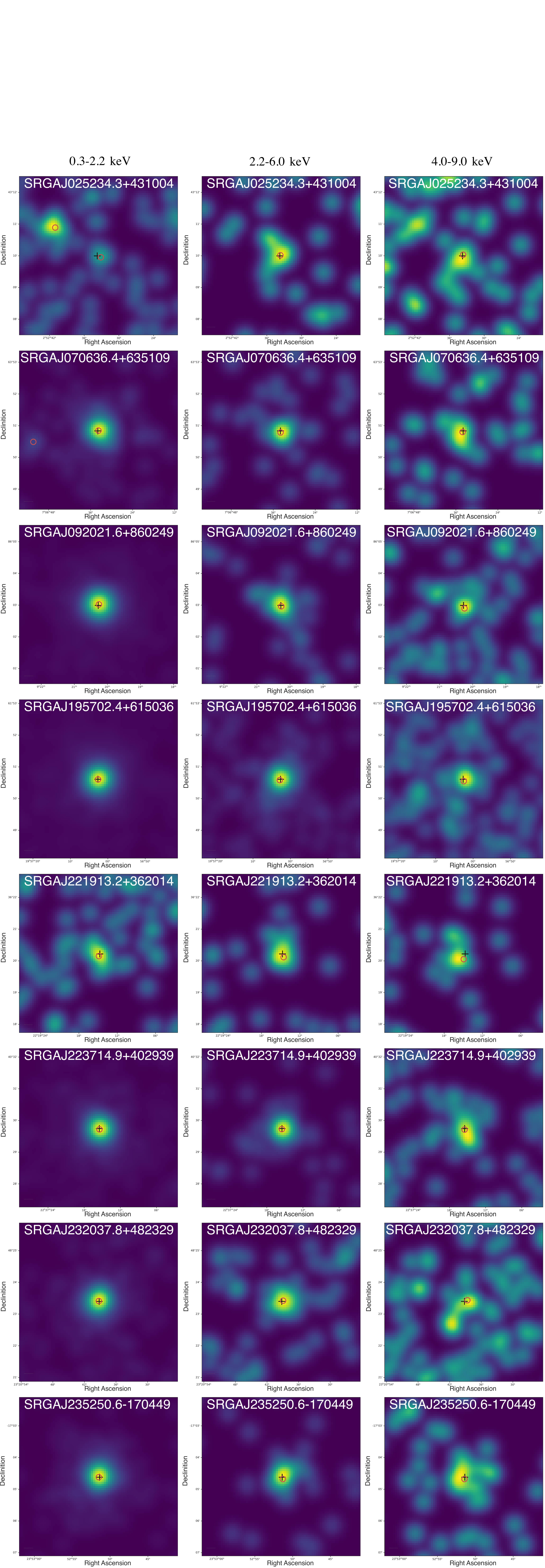}
    \caption{
    \ero\ images for the eight sources located on the half of the sky processed by the Russian \ero\ consortium in the 0.3--2.2 (left), 2.2--6.0 (center), and 4.0--9.0 (right) keV energy bands. The image size is 5x5 arcmin. The circles mark the positions of the sources detected by \ero\ at a confidence level higher than $3\sigma$. The \art\ positions of the sources being discussed here are indicated by the black cross
    }
    \label{fig:ero}
\end{figure}

Among the 17 sources from the \art\ catalog selected for our studies, eight are located on the half of the sky on which the Russian \ero\ consortium is responsible for processing the \ero\ data. All of the sources are detected by the \ero\ telescope in both soft (0.3--2.2 keV) and hard (4--9 keV) energy bands, except for SRGA J025234.3+431004 that is reliably ($\ga 5\sigma$) detected only at energies $\ga 2.2$ keV. The \ero\ images of all eight sources in the 0.3--2.2, 2.2--6.0, and 4.0--9.0 keV energy bands are presented in Fig.~\ref{fig:ero}.

Our spectral analysis was performed jointly using the \art\ and \ero\ data. 
The spectra were fitted in the energy range 0.2--20~keV with the XSPEC v12.11.0n\footnote{https://heasarc.gsfc.nasa.gov/xanadu/xspec/} software \citep{arnaud1996}. 
The W-statistic, a modified C-statistic \citep{cash1979}, in which the Poisson background around the source is taken into account was used for model fitting.
 
A preliminary version of the response matrix, which was prepared based on the observations of the Crab Nebula and its pulsar, was used in our spectral analysis of the \art\ data. To investigate the relative calibration of the \art\ and \ero\ telescopes, we chose comparatively bright sources from our sample and fitted simultaneously their spectra by power laws with absorptions. At the same time, a cross-calibration constant was added. It was fixed at unity for \ero\ and assumed to be the same for all sources for \art. We obtained a value of $1.3$ for this constant and the corresponding 90\% confidence interval ($1.0$--$1.7$). Since the value of the cross-calibration constant is compatible with unity, we concluded that its introduction is not required. Note that the work to refine the \art\ response matrix continues, and a value of the cross-calibration constant different from unity may be required in analyzing the data of succeeding releases.

The X-ray radiation from AGNs can experience absorption in the gas–dust torus around the supermassive black hole (SMBH) and in the interstellar medium of the host galaxy. One of the goals of our study was to estimate the gas column density $\nh$ inside the objects being investigated. To describe the X-ray spectra, we used the standard (for AGN research) model of a power-law continuum with a low- energy cutoff as a result of photoabsorption in the Galaxy and the object itself. At energies below 2 keV an excess of X-ray emission (soft excess), whose nature is debated (see, e.g., \cite{boissay2016}), is often observed in the X-ray spectra of AGNs. Given the comparatively small number of photons in the spectra being studied here, we used a simple phenomenological model of blackbody radiation with a fixed temperature $kT=0.1$~keV to describe the soft component in the spectrum. Thus, we used two models in \textsc{XSPEC}:
\[
phabs(zphabs(cflux\,powerlaw)),
\]
\[
phabs(zphabs(cflux\,powerlaw) + blackbody),
\]
where $phabs$ is the absorption in the Galaxy from HI4PI data \citep{hi4pi:a2016}, zphabs is the absorption at the redshift $z$ of a given AGN (measured from the object’s optical spectrum), $cflux$ is the convolution model that normalizes the power-law component in flux in the 4--12~keV energy band (the absorption-corrected flux is determined in this way).

To compare the qualities of the fits between these two models, we used the Akaike information criterion \citep{akaike1974} $AIC=2k+cstat$, where $k$ is the number of free model parameters and cstat is the value of the likelihood function -2log$L_{\rm max}$ \citep{cash1979}. If AIC decreased by 5 or more when adding the blackbody component (this corresponds to the fact that the realization probability of the first model is no more than 8\% of the realization probability of the second one), then preference was given to the two-component model.

The spectral fitting results are presented in Table \ref{tab:xray_params}. The 90\% confidence intervals of the parameters are given. The X-ray spectra themselves are presented below in the subsections devoted to the individual sources (Figs. \ref{fig:spec0252}--\ref{fig:spec2352}). Some of the \ero\ spectra were rebinned for better visual perception. For three sources (SRGA J025234.3+431004, SRGA J221913.2+362014, SRGA J235250.6-170449), in which our spectral analysis revealed an intrinsic absorption, we also show the two-dimensional regions of spectral slopes $\Gamma$ and absorption column densities.

When analyzing Figs. \ref{fig:spec0252}--\ref{fig:spec2352}, it can be noticed that the observed flux in the first \art\ energy channel in the spectra for seven of the eight sources exceeds the value predicted by the model. In some cases, this excess is considerable and statistically significant. Part of the observed discrepancy between the \art\ data and the model is probably associated with the Eddington bias, which stems from the fact that the sources have a low significance in the \art\ data and are near the detection threshold. The drawbacks of the current \art\ calibration, which will be removed in succeeding data releases, can also contribute noticeably to the observed discrepancy. In the context of interpreting the results of our joint spectral analysis of the \ero\ and \art\ data presented here, it should be emphasized that the integral response of the \art\ telescope in the 4--20~keV energy band was calibrated based on the Crab observations reasonably well and reproduces its flux with an accuracy $\sim 20\%$, which allows these data to be used to analyze the broadband spectra. We should also take into account the fact that for most of the sources the spectral parameters are largely determined by the \ero\ data with a higher statistical significance.

\begin{table*}
    \caption{X-ray spectral parameters}
    \label{tab:xray_params}
    \vskip 2mm
    \renewcommand{\arraystretch}{1.1}
    \renewcommand{\tabcolsep}{0.35cm}
    \centering
    \footnotesize
\begin{tabular}[t]{lcccccrr}
\noalign{\doubleline}
\multispan7\hfil PHA(ZPHA\,CFLUX\,PL) \hfil \\
$\art$ source & $N_{\rm H}^{MW}$ & $\nh$ & $\Gamma$ & $F_{\rm PL}^{4-12}$ & $A_{\rm BB}$, $10^{-6}$ & Cstat (dof) & AIC \\
\noalign{\vskip 3pt\hrule\vskip 5pt}
SRGA\,J$025234.3\!+\!431004$ & $1.0$ & $78^{+391}_{-71}$ & $0.6^{+2.4}_{-1.3}$               & $2.4^{+4.9}_{-1.2}$ & $-$ & $12$ ($10$)   & $31.6$ \\[0.3em]
SRGA\,J$070636.4\!+\!635109$ & $0.4$ & $<0.2$                  & $1.1^{+0.2}_{-0.2}$ & $1.9^{+0.9}_{-0.7}$  & $-$ & $68$ ($53$)   & $173.8$\\[0.3em]
SRGA\,J$092021.6\!+\!860249$ & $0.5$ & $<0.2$                  & $2.2^{+0.1}_{-0.1}$ & $1.0^{+0.2}_{-0.2}$  & $-$ & $230$ ($212$) & $654$  \\[0.3em]
SRGA\,J$195702.4\!+\!615036$ & $0.7$ & $<0.3$                  & $1.7^{+0.1}_{-0.1}$ & $2.3^{+0.5}_{-0.4}$  & $-$ & $289$ ($283$) & $854.9$\\[0.3em]
SRGA\,J$221913.2\!+\!362014$ & $1.0$ & $80^{+53}_{-41}$ & $1.8^{+1.0}_{-0.9}$ & $2.2^{+1.2}_{-0.9}$    & $-$ & $19$ ($14$)   & $47$   \\[0.3em]
SRGA\,J$223714.9\!+\!402939$ & $1.2$ & $<0.6$                  & $1.4^{+0.1}_{-0.1}$ & $5.2^{+1.1}_{-0.9}$  & $-$ & $183$ ($189$) & $560.9$\\[0.3em]
SRGA\,J$232037.8\!+\!482329$ & $1.3$ & $<0.4$                  & $1.5^{+0.2}_{-0.2}$ & $1.4^{+0.5}_{-0.4}$  & $-$ & $119$ ($92$)  & $302.5$\\[0.3em]
SRGA\,J$235250.6\!-\!170449$ & $0.2$ & $1.7^{+0.7}_{-0.6}$     & $1.3^{+0.2}_{-0.2}$ & $6.0^{+2.1}_{-1.7}$  & $-$ & $94$ ($120$)  & $333.7$\\[0.3em]
\noalign{\vskip 3pt\hrule\vskip 5pt}
\multispan7\hfil PHA(ZPHA\,CFLUX\,PL + BB) \hfil \\
\noalign{\vskip 3pt\hrule\vskip 5pt}
SRGAJ070636.4+635109 & $0.4$ & $<4.8$ & $1.0^{+0.5}_{-0.4}$ & $2.3^{+1.5}_{-1}$                 & $ 3.1^{+1.4}_{-2.1}$ & $61$ ($52$)   & $165.4$\\[0.3em]
SRGAJ223714.9+402939 & $1.2$ & $ 2.3^{+1.6}_{-1.5}$ & $1.7^{+0.3}_{-0.3}$ & $4.1^{+1.5}_{-1.2}$ & $14.1^{+5.6}_{-7.4}$ & $177$ ($189$) & $554.5$\\[0.3em]

\hline
\end{tabular}
  \begin{flushleft}
  $N_{\rm H}^{MW}$ and $\nh$ are the gas column densities in the Milky Way Galaxy and the object, respectively, in units of $10^{21}$ cm$^{-2}$;  
  $F_{\rm PL}^{4-12}$ is the absorption-corrected flux in the 4--12~keV energy band created by the power-law component, in units of $10^{-12}$~erg~s$^{-1}$~cm$^{-2}$;  
  $A_{\rm BB}$ is the normalization of the blackbody component.
  \end{flushleft}
\end{table*}

\section{OPTICAL OBSERVATIONS}

\begin{table*}
    \caption{Log of optical observations}
    \label{tab:list_obs}
    \vskip 2mm
    \renewcommand{\arraystretch}{1.1}
    \renewcommand{\tabcolsep}{0.35cm}
    \centering
    \footnotesize
\begin{tabular}[t]{lccccl}
\noalign{\doubleline}
\art\ source & Date & Grism & Slit & Exposure time, s & Telescope\\
\noalign{\vskip 3pt\hrule\vskip 5pt}
SRGA\,J$025234.3\!+\!431004$ & 2021-09-29 & VPHG600G & $2\arcsec$ & $3\times300$ & \azt\\
SRGA\,J$062627.2\!+\!072734$ & 2021-11-05 & G15 & $2\arcsec$ & $8\times600$ & \rtt\\
SRGA\,J$070636.4\!+\!635109$ & 2021-05-13 & VPHG600G & $2\arcsec$ & $7\times200$ & \azt\\
SRGA\,J$092021.6\!+\!860249$ & 2021-10-31 & VPHG600G & $2\arcsec$ & $4\times300$ & \azt\\
SRGA\,J$195702.4\!+\!615036$ & 2021-05-12 & G15 & $2\arcsec$ & $5\times600$ & \rtt\\
SRGA\,J$221913.2\!+\!362014$& 2021-10-31 & VPHG600G & $3\arcsec$ & $3\times600$ & \azt\\
& 2021-10-31 & VPHG600R & $3\arcsec$ & $3\times600$ & \azt\\
SRGA\,J$223714.9\!+\!402939$ & 2021-05-13 & VPHG600G & $3\arcsec$ & $3\times200$ & \azt\\
SRGA\,J$232037.8\!+\!482329$ & 2021-11-05 & G15 & $2\arcsec$ & $4\times120$ & \rtt\\
SRGA\,J$235250.6\!-\!170449$ & 2021-09-11 & VPHG600G & $2\arcsec$ & $4\times300$ & \azt\\
\noalign{\vskip 3pt\hrule\vskip 5pt}
\end{tabular}
\end{table*}

Our spectroscopy for the northern-sky ($\delta>-20^\circ$) objects was performed at the RTT-150 telescope using the \emph{TFOSC}\footnote{http://hea.iki.rssi.ru/rtt150/en/index.php?page=tfosc} spectrograph and at the AZT-33IK telescope using the low- and medium-resolution ADAM spectrograph \citep{afanasiev2016, burenin2016}(see the log of observations in Table~\ref{tab:list_obs}). We used long slits of width 2\arcsec\ and 3\arcsec\ at the ADAM spectrograph and of 2\arcsec at the \emph{TFOSC} spectrograph. The slit center was brought into coincidence with the central region of the observed galaxy. After each exposure, the object’s position was shifted along the slit by 10--15$\arcsec$ in a random direction upward or down- ward using a photoguide. The optical observations were performed at a seeing better than 2.5\arcsec.

Transmitting diffraction grating no. 15 with the spectral range 3700--8700~\AA\ providing a spectral resolution of 12~\AA was used at the \emph{TFOSC} spectrograph as a dispersive element. This grating allows bright Balmer lines to be obtained in the spectral images for galaxies up to $z=0.32$. The spectrograph slit position angle is 90$\circ$. Before and after obtaining the series of spectroscopic images for each object, we obtained the images of a lamp with a continuum spectrum and the line spectrum of a Fe–Ar lamp.

We used volume phase holographic gratings (VPHG), 600 lines per millimeter, to take the spectra at the ADAM spectrograph. As a dispersive element we used VPHG600G for the spectral range 3650--7250~\AA\ with a resolution of 8.6~\AA\ for a 2\arcsec slit and 12.9~\AA\ for a 3\arcsec\ slit and VPHG600R for the spectral
range 6460--10\,050~\AA\ with a resolution of 18.3~\AA\ for a 3\arcsec slit. When using VPHG600R, we set the
OS11 filter, which removes the second interference order from the image. A thick e2v CCD30-11 array produced by the deep depletion technology is installed at the spectrograph. This allows the spectral images
to be obtained at a wavelength of 10\,000\AA\ without interference on the thin CCD substrate. All our observations were performed with zero slit position angle. After each series of spectroscopic images for each object, we obtained the calibration images of a lamp with a continuum spectrum and the line spectrum of a He–Ne–Ar lamp.

On each observing night we took the spectra of spectrophotometric standards from the ESO\footnote{https://www.eso.org/sci/observing/tools/standards} list for all the sets of diffraction gratings and slits being used. The spectrophotometric standards were chosen so that they were approximately at the same elevation with the optical source observed by us. The data reduction was performed using the IRAF5 software and our own software. The flux calibration was performed by standard IRAF\footnote{http://iraf.noao.edu/} procedures from the onedspec package.

\section{RESULTS OF OBSERVATIONS}

Here we discuss the results of our observations of the northern-sky objects. 
The emission lines were fitted by a Gaussian to determine such parameters as the line center, full width at half maximum $FWHM_{\rm mes}$, flux, and equivalent width $EW$. 
The spectral continuum was fitted by a polynomial whose order depended on the spectral shape. The FWHM of the Balmer lines was corrected for the spectral resolution of the instrument: ${FWHM} = \sqrt{{FWHM}^2_{\rm mes} - {FWHM}^2_{\rm res}}$ ,where ${FWHM}_{\rm res}$ was determined for each dispersive element and each slit as the FWHM of the lines in the calibration lamp spectrum. 
The FWHMs of the narrow lines are consistent with the instrumental broadening ${FWHM}_{\rm res}$ and, therefore, the values of $FWHM$ are not given for them in Tables \ref{tab:j0252}--\ref{tab:j2352}.

Standard criteria \citep{osterbrock1981, veron-cetty2001} were used to classify the Seyfert galaxies. The measurement errors of the emission line parameters are given at the 68\% confidence level.

The confidence level of the redshift was determined as the error of the mean of the narrow-line redshifts.

\subsection{\it SRGA J025234.3+431004}

This source was discovered in the \swift/BAT hard
X-ray sky survey (PBC J0252.3+4309 = SWIFT J0252.3+4312) \citep{cusumano2010, oh2018}. There is the edge-on galaxy LEDA 90641 with the infrared color $W1-W2=0.77$ that points to the probable presence of an active galactic nucleus in the \art\ position error circle. According to the SIMBAD astronomical database, the galaxy’s redshift is $z= 0.0518$.

\begin{figure*}
  \centering
  \vfill
  SRGA\,J$025234.3\!+\!431004$
  \vfill
  \vskip 0.5cm
  \begin{floatrow}
    \includegraphics[width=0.3\columnwidth]{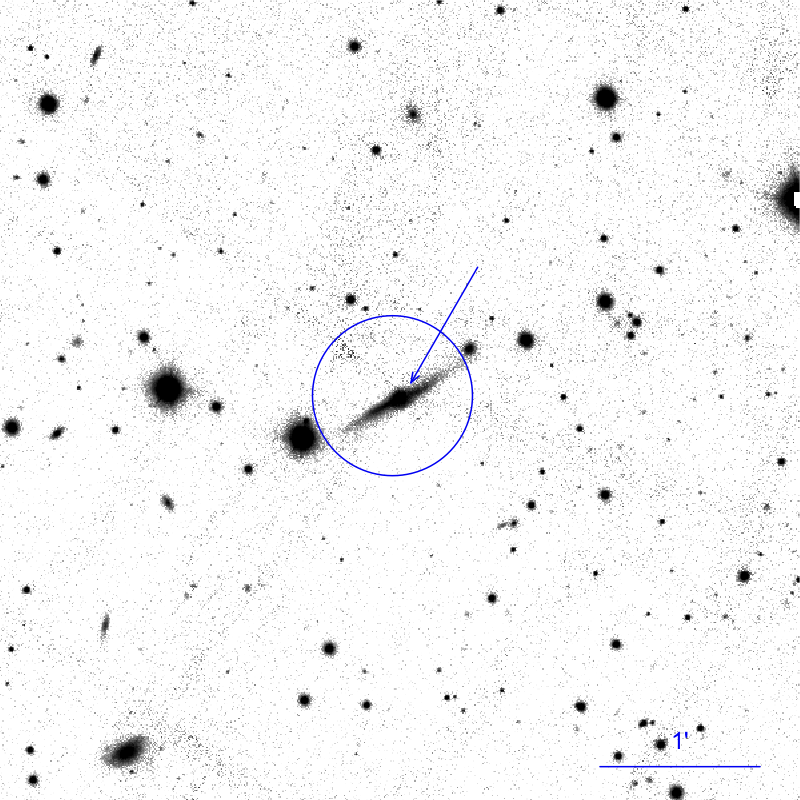}
   \includegraphics[width=0.4\columnwidth]{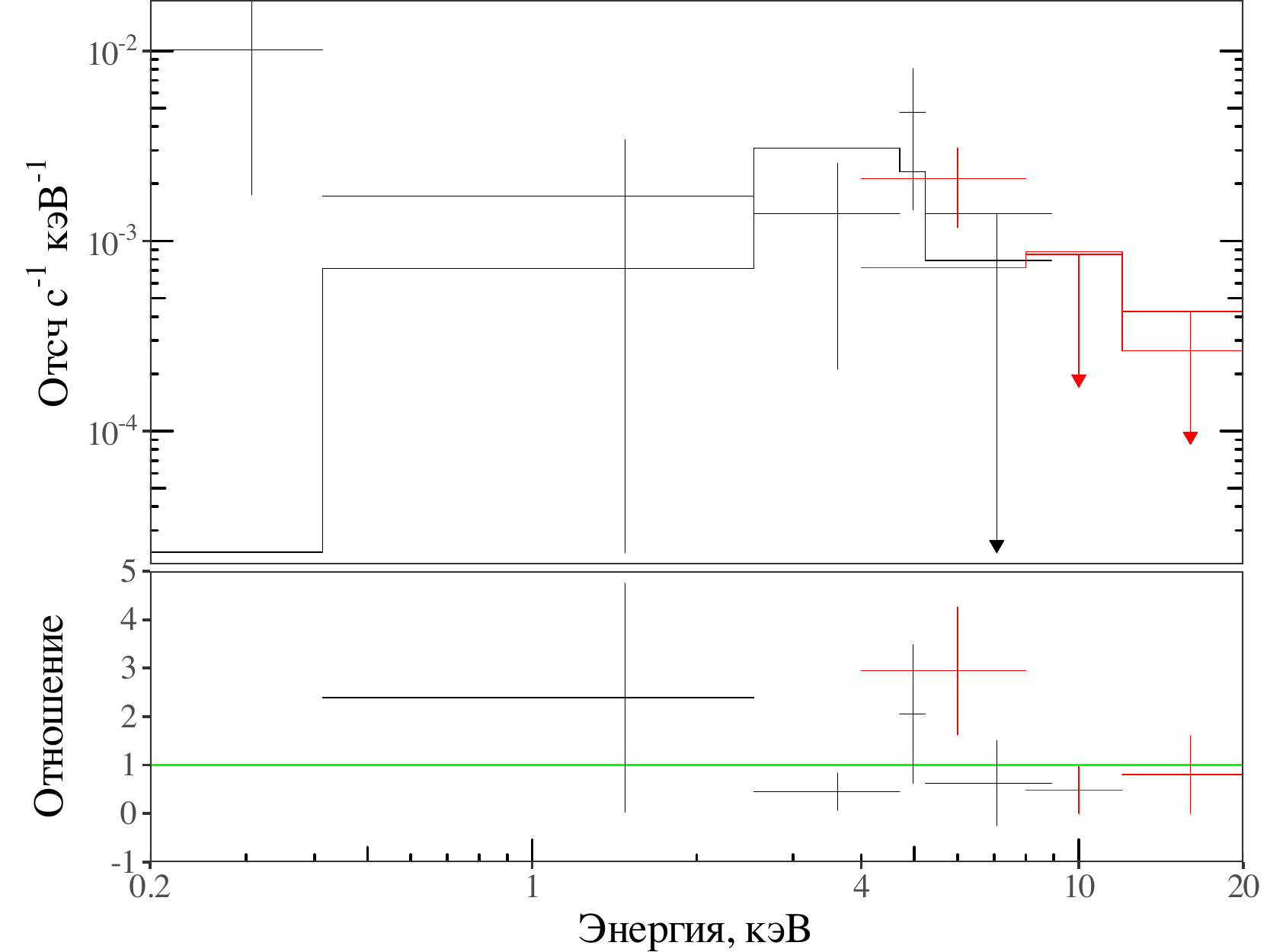}
    \includegraphics[width=0.35\columnwidth]{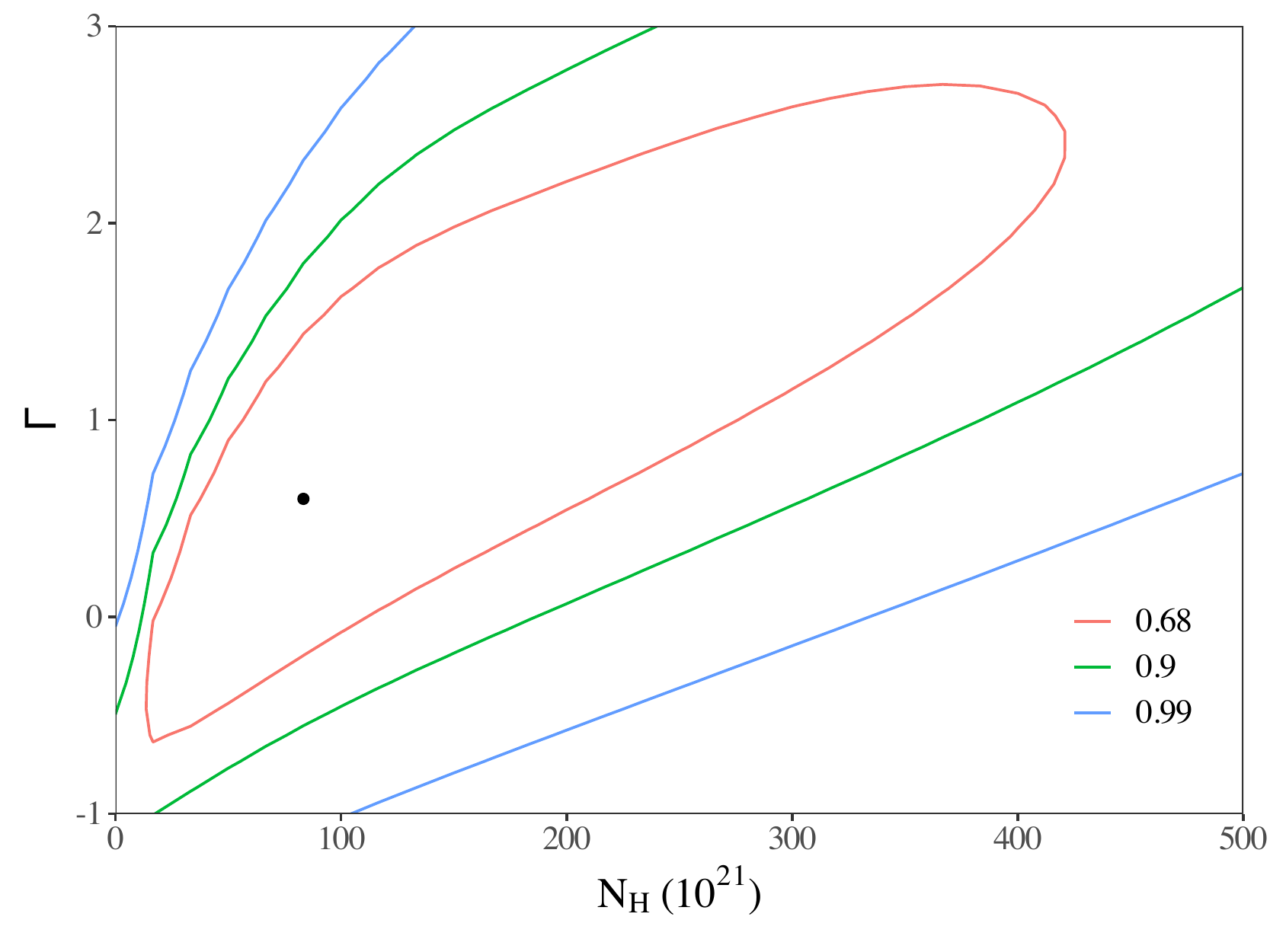}
  \end{floatrow}
  \vfill
 \vspace{1cm}
  \vfill
  \vspace{-1cm}
  \begin{floatrow}
    \includegraphics[width=0.6\columnwidth]{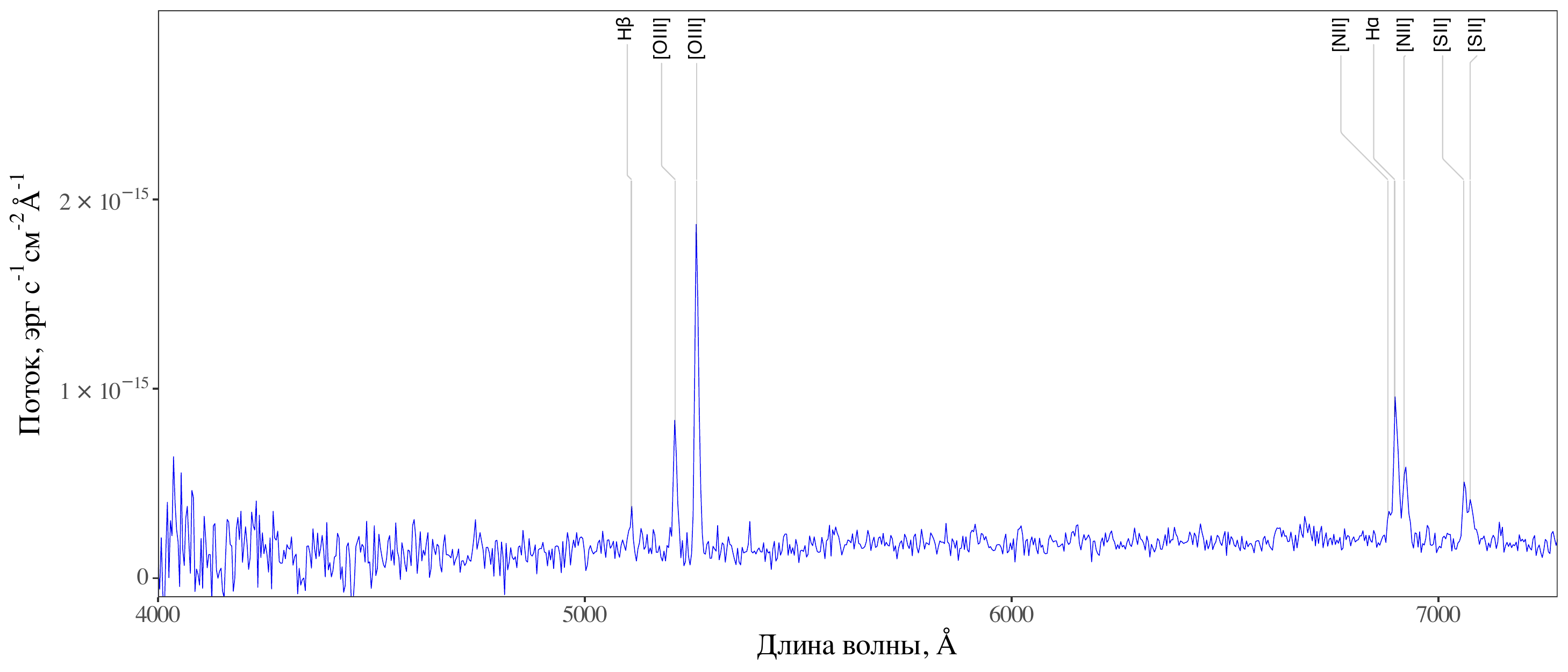}

  \end{floatrow}
  \caption{
  Results of the observations of SRGA\,J$025234.3\!+\!431004$. Top left: the pointing picture. Only the \art\ position error circle is indicated by the blue circumference, with a radius of $30\arcsec$, due to an insufficiently high detection significance of the source by the \ero\ telescope. The arrow indicates the object for which an optical spectrum was taken. Top middle: the X-ray spectrum from the \art\ (red) and \ero\ (black) data and the best-fit model (see Table~\ref{tab:xray_params}). The arrows indicate the 2 $\sigma$ upper limits. The measurement-to-model ratio is shown on the lower panel of the X-ray spectrum. The cstat likelihood contours of the best-fit model of the X-ray spectrum for the slope $\Gamma$ and the column density $\nh$ are shown to the right of the X-ray spectrum. The inner red, green, and outer blue contours correspond to the 68\%, 90\%, and 99\% confidence intervals, respectively. Bottom: the optical spectrum, the main emission lines are indicated. 
  }
  \label{fig:spec0252}
\end{figure*}

The galaxy’s spectrum (Fig.~\ref{fig:spec0252}, Table~\ref{tab:j0252}) exhibits narrow H$\beta$, 
$[OIII]\lambda$4959,  
$[OIII]\lambda$5007, 
$[NII]\lambda$6548, 
H$\alpha$, 
$[NII]\lambda$6584, 
and sulfur doublet lines, 
from which the redshift can be refined: $z = 0.05123\pm 0.00024$. 
By its position on the BPT diagram (lg([OIII]$\lambda 5007/$H${\beta})=1.01\pm0.11$ 
and lg([NII]$\lambda 6584/$H${\alpha})=-0.24\pm0.04$), 
the object can be attributed to Seyfert galaxies, while the absence of broad line components implies that this is Sy2.

\begin{table*}
  \caption{Spectral features of SRGA\,J$025234.3\!+\!431004$} 
  \label{tab:j0252}
  \vskip 2mm
  \renewcommand{\arraystretch}{1.1}
  \renewcommand{\tabcolsep}{0.35cm}
  \centering
  \footnotesize
  \begin{tabular}{lcccc}
    \noalign{\doubleline}
    Line & Wavelength, \AA & Flux, $10^{-15}$~erg~s$^{-1}$~cm$^{-2}$ & Eq. width, \AA & $FWHM$, $10^2$ km/s\\
    \noalign{\vskip 3pt\hrule\vskip 5pt}
H$\beta$              & 5109 & $1.8\pm0.5$ & $- 11.5\pm3.0$ & $-$\\
{}[OIII]$\lambda$4959 & 5212 & $6.8\pm0.5$ & $- 44\pm3$ & $-$\\
{}[OIII]$\lambda$5007 & 5262 & $17.8\pm0.1$& $-117\pm4$ & $-$\\
{}[NII]$\lambda$6548  & 6884 & $1.2\pm0.3$ & $-  5.9\pm1.4$ & $-$\\
H$\alpha$             & 6900 & $9.3\pm0.4$ & $- 46\pm2$ & $-$\\
{}[NII]$\lambda$6584  & 6923 & $5.3\pm0.4$ & $- 27\pm2$ & $-$\\
{}[SII]$\lambda$6718  & 7061 & $2.6\pm0.5$ & $- 13.2\pm2.6$ & $-$\\
{}[SII]$\lambda$6732  & 7075 & $3.2\pm0.6$ & $- 16.0\pm3.1$ & $-$\\
    \noalign{\vskip 3pt\hrule\vskip 5pt}
  \end{tabular}
\end{table*}

\subsection{\it SRGA J062627.2+072734}

\begin{figure*}
  \centering
  \vfill
  SRGA\,J$062627.2\!+\!072734$
  \vfill
  \vskip 0.5cm
  \begin{floatrow}
    \includegraphics[width=0.3\columnwidth]{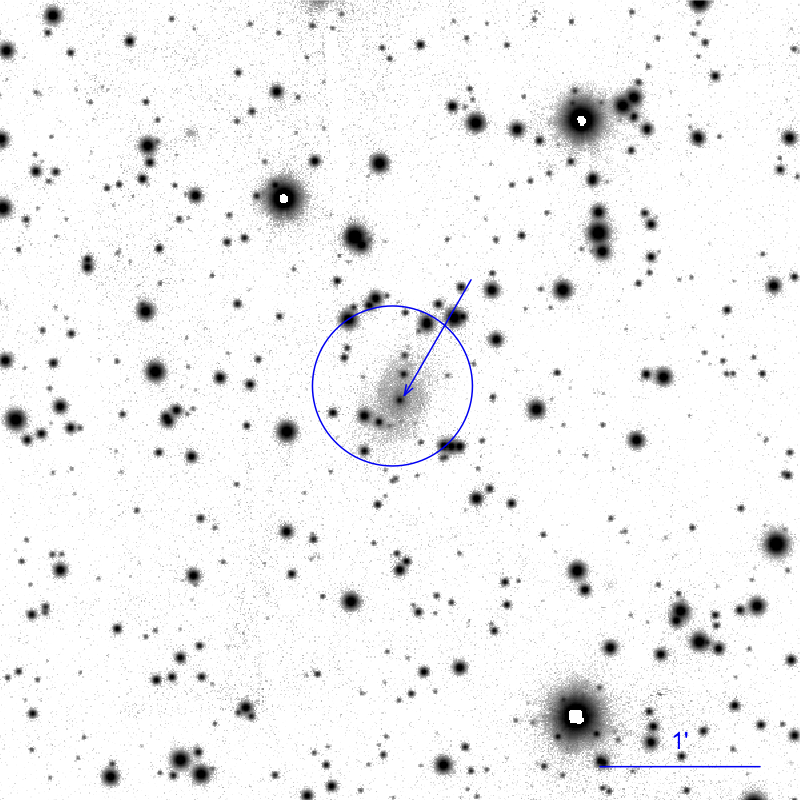}
  \end{floatrow}
  \vfill
  \vspace{1cm}
  \vfill
  \vspace{-1cm}
  \begin{floatrow}
  \includegraphics[width=0.6\columnwidth]{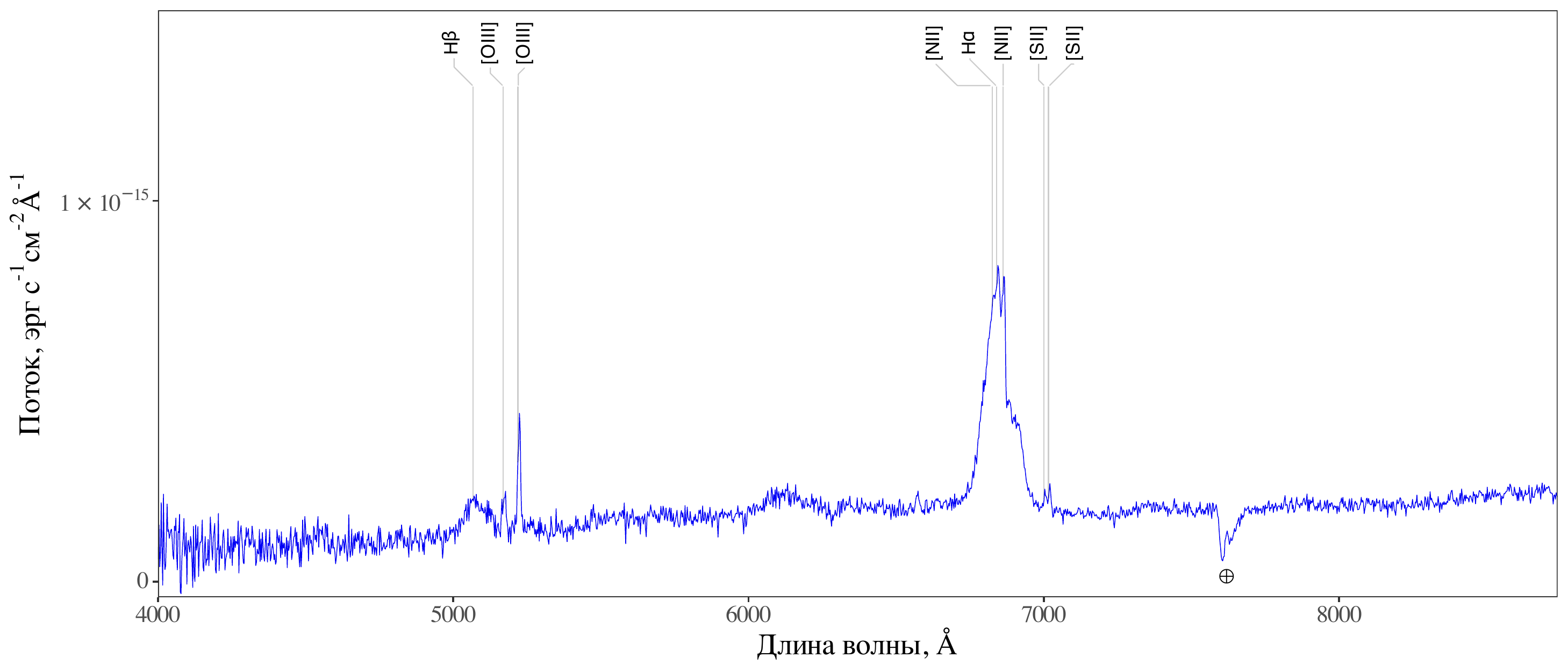}
  \end{floatrow}
  \caption{
  Same as Fig.~\ref{fig:spec0252}, but for SRGA\,J$062627.2\!+\!072734$. The \ero\ data (the position error circle, the X-ray spectrum, and the cstat contours) are not shown due to the source location on the half of the sky $180<|l|<360^\circ$.
  }
  \label{fig:spec0626}
\end{figure*}

This source was discovered during the ROSAT all-sky survey (2RXS J062625.8+072733). It is also present in the catalog of sources of the \swift/BAT hard X-ray survey (SWIFT J0626.6+0729) \citep{oh2018}. The object is on the half of the sky $180<|l|<360^\circ$, on which we have no \ero\ data at our disposal. There is the galaxy LEDA\,136513 (Fig.~\ref{fig:spec0626}) with an infrared color typical for AGNs ($W1-W2=0.86$) in the ART-XT position error circle.

The galaxy’s optical spectrum (Fig.~\ref{fig:spec0626}, Table~\ref{tab:j0626}) exhibits the H$\alpha$ and H$\beta$ emission lines with intense broad components as well as narrow forbidden [OIII]$\lambda$4959, [OIII]$\lambda$5007, [NII]$\lambda$6548, [NII]$\lambda$6584, and sulfur doublet lines. The redshift is determined from these lines: $z = 0.04254\pm0.00013$. The ratios lg([OIII]$\lambda 5007/$H${\beta})>1.01$ and lg([NII]$\lambda 6584/$H${\alpha})=-0.03\pm0.08$ measured for the narrow H$\alpha$ and H$\beta$ components are typical for AGNs, while the presence
of broad H$\alpha$ and H$\beta$ components with fluxes higher than those in the narrow components by an order of magnitude allows this object to be attributed to Seyfert 1 galaxies (Sy1).

\begin{table*}
  \caption{Spectral features of SRGA\,J$062627.2\!+\!072734$} 
  \label{tab:j0626}
  \vskip 2mm
  \renewcommand{\arraystretch}{1.1}
  \renewcommand{\tabcolsep}{0.35cm}
  \centering
  \footnotesize
  \begin{tabular}{lcccc}
    \noalign{\doubleline}
    Line & Wavelength, \AA & Flux, $10^{-15}$~erg~s$^{-1}$~cm$^{-2}$ & Eq. width, \AA & $FWHM$, $10^2$ km/s\\
    \noalign{\vskip 3pt\hrule\vskip 5pt}
H$\beta$, narrow      & 5077 & $<0.3       $ & $>-2.5$        & $-$                      \\
H$\beta$, broad       & 5077 & $7.9\pm 0.5 $ & $- 64\pm4$ & $54\pm4$\\
{}[OIII]$\lambda$4959 & 5173 & $1.3\pm 0.2 $ & $- 10.1\pm1.3$ & $-$\\
{}[OIII]$\lambda$5007 & 5224 & $3.2\pm 0.2 $ & $- 24\pm1$ & $-$\\
{}[NII]$\lambda$6548  & 6826 & $1.9\pm 0.3 $ & $- 10.4\pm1.6$ & $-$\\
H$\alpha$, broad    & 6841 & $55\pm 1$ & $-297\pm5$ & $53\pm1$\\
H$\alpha$, narrow      & 6841 & $2.5\pm 0.3 $ & $- 13.2\pm1.6$ & $-$\\
{}[NII]$\lambda$6584  & 6859 & $2.3\pm 0.3 $ & $- 12.2\pm1.6$ & $-$\\
{}[SII]$\lambda$6718  & 7002 & $0.5\pm 0.2 $ & $-  2.7\pm1.0$ & $-$\\
{}[SII]$\lambda$6732  & 7017 & $0.5\pm 0.2 $ & $-  2.8\pm0.8$ & $-$\\
    \noalign{\vskip 3pt\hrule\vskip 5pt}
  \end{tabular}
\end{table*}

\subsection{\it SRGA J070636.4+635109}

\begin{figure*}
  \centering
  \vfill
  SRGA\,J$070636.4\!+\!635109$
  \vfill
  \vskip 0.5cm
  \begin{floatrow}
    \includegraphics[width=0.3\columnwidth]{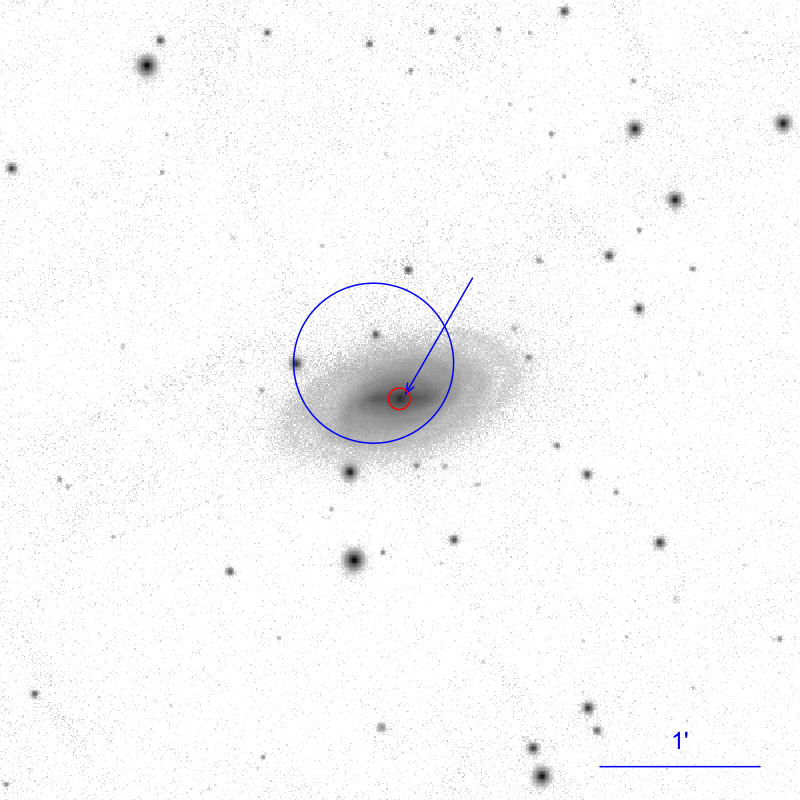}
   \includegraphics[width=0.4\columnwidth]{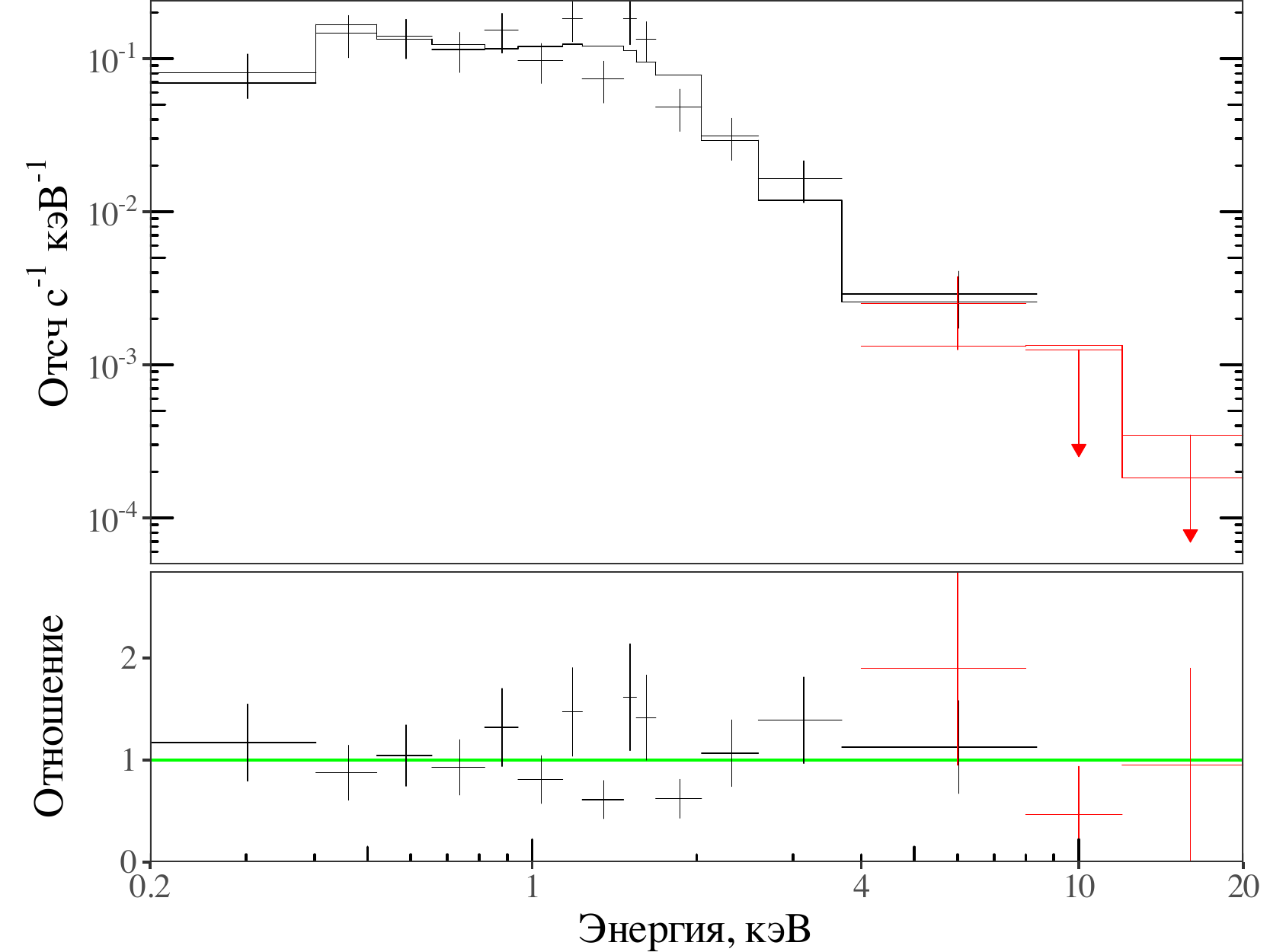}
  \end{floatrow}
  \vfill
 \vspace{1cm}
  \vfill
  \vspace{-1cm}
  \begin{floatrow}
  \includegraphics[width=0.6\columnwidth]{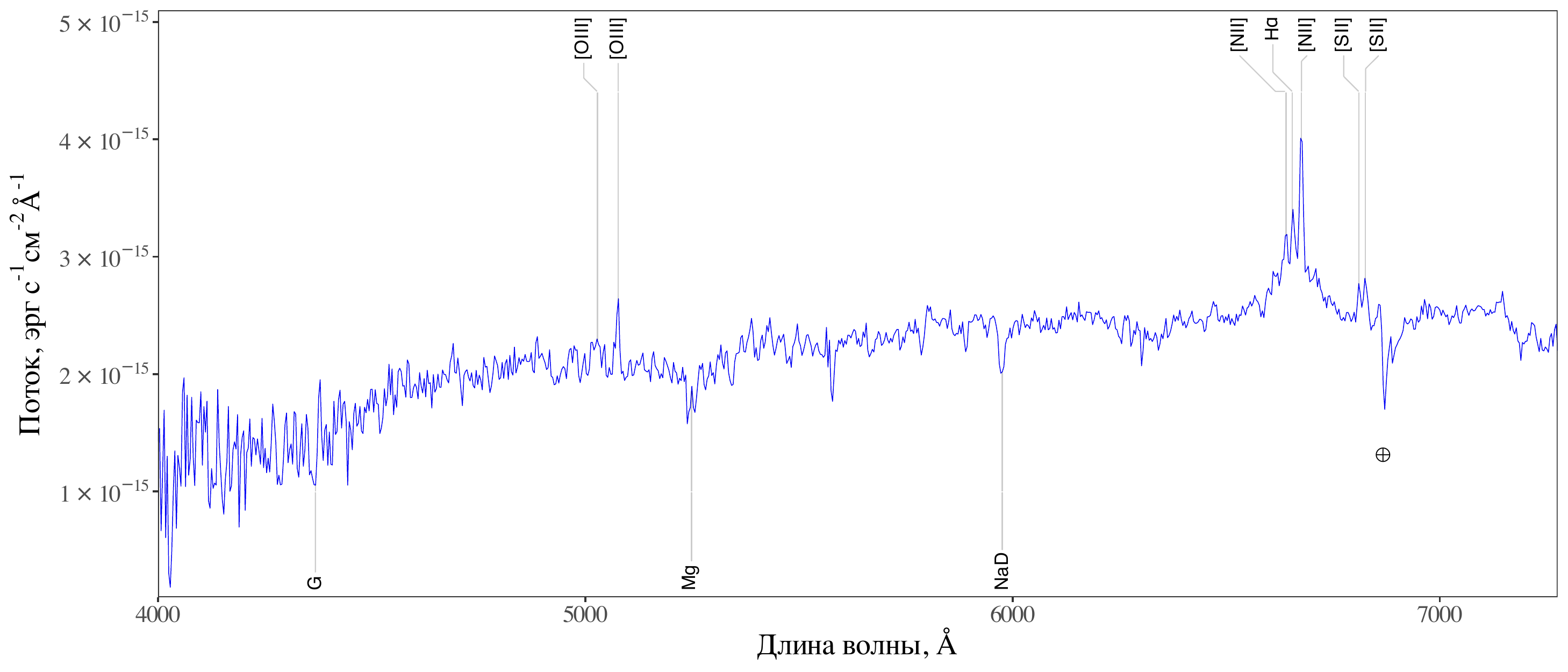}
  \end{floatrow}
  \caption{
  Results of the observations of SRGA\,J$070636.4\!+\!635109$. Top left: the pointing picture. The blue circumference indicates the \art\ source position error circle, with a radius of $30\arcsec$; the red circumference indicates the 98\% \ero\ position error circle. The arrow indicates the object for which an optical spectrum was taken. Top right: the X-ray spectrum from the \art\ (red) and \ero\ (black) data and the best-fit model (see Table~\ref{tab:xray_params}). The arrows indicate the 2 $\sigma$ upper limits. The measurement-to-model ratio is shown on the lower panel of the X-ray spectrum. Bottom: the optical spectrum, the main emission lines are indicated.
  }
  \label{fig:spec0706}
\end{figure*}

This X-ray source was discovered by the \art\ telescope onboard the \srg\ observatory during the first year of its all-sky survey \citep{pavlinsky2022}. It was also detected by the \integral\ gamma-ray observatory during a long-term hard X-ray all-sky survey (for a review, see \cite{krivonos2021}) on the maps of the galaxy M81 (Mereminskiy et al., in preparation) with a flux of $(8.6\pm2.0)\times10^{-12}$ erg s$^{-1}$ cm$^{-2}$ in the 17--60~keV energy band. There is the galaxy UGC~3660 at $z=0.0143$ (according to SIMBAD) in the \art\ position error circle refined based on the \ero\ data (Fig.~\ref{fig:spec0706}) with which the radio source NVSS J070632+635101 \citep{condon1998} can also be associated.

The galaxy’s spectrum (Fig.~\ref{fig:spec0706}, Table~\ref{tab:j0706}) exhibits narrow [OIII]$\lambda$4959, [OIII]$\lambda$5007, [NII]$\lambda$6548, H${\alpha}$, [NII]$\lambda$6584, and sulfur doublet emission lines; a broad H${\alpha}$ component is seen. At the same time, the H${\beta}$ line is not detected. The spectrum also exhibits the G, MgI, and NaD absorption lines of the Fraunhofer series. The object’s redshift measured from the lines is $z=0.01404\pm 0.00019$.

By its position on the BPT diagram (lg([OIII]$\lambda 5007/$H${\beta})>0.65$, 
lg([NII]$\lambda 6584/$H${\alpha})=0.42\pm0.07$), the object can be attributed to Seyfert or LINER galaxies. However, the presence of a broad H${\alpha}$ component and the absence of H${\beta}$ allow the object to be classified as a Seyfert 1.8 galaxy (Sy1.8).

\begin{table*}
  \caption{Spectral features of SRGA\,J$070636.4\!+\!635109$} 
  \label{tab:j0706}
  \vskip 2mm
  \renewcommand{\arraystretch}{1.1}
  \renewcommand{\tabcolsep}{0.35cm}
  \centering
  \footnotesize
  \begin{tabular}{lcccc}
    \noalign{\doubleline}
    Line & Wavelength, \AA & Flux, $10^{-15}$~erg~s$^{-1}$~cm$^{-2}$ & Eq. width, \AA & $FWHM$, $10^2$ km/s\\
    \noalign{\vskip 3pt\hrule\vskip 5pt}
H$\beta$              & 4930 & $<1.1$       & $>-0.5$       & $-$\\
{}[OIII]$\lambda$4959 & 5027 & $2.4\pm0.8 $ & $- 1.2\pm0.4$ & $-$\\
{}[OIII]$\lambda$5007 & 5076 & $4.9\pm0.9 $ & $- 2.4\pm0.4$ & $-$\\
{}[NII]$\lambda$6548  & 6640 & $2.5\pm0.6 $ & $- 1.0\pm0.2$ & $-$\\
H$\alpha$             & 6657 & $4.4\pm0.7 $ & $- 1.8\pm0.3$ & $-$\\
H$\alpha$, broad      & 6657 & $68\pm5$     & $-28\pm2$     & $61\pm4$\\
{}[NII]$\lambda$6584  & 6676 & $11.4\pm0.1$ & $- 4.7\pm0.3$ & $-$\\
{}[SII]$\lambda$6718  & 6811 & $2.5\pm0.5 $ & $- 1.0\pm0.2$ & $-$\\
{}[SII]$\lambda$6732  & 6826 & $4.0\pm0.7 $ & $- 1.6\pm0.3$ & $-$\\
    \noalign{\vskip 3pt\hrule\vskip 5pt}
  \end{tabular}
\end{table*}

\subsection{\it SRGA J092021.6+860249}

\begin{figure*}
  \centering
  \vfill
  SRGA\,J$092021.6\!+\!860249$
  \vfill
  \vskip 0.5cm
  \begin{floatrow}
    \includegraphics[width=0.3\columnwidth]{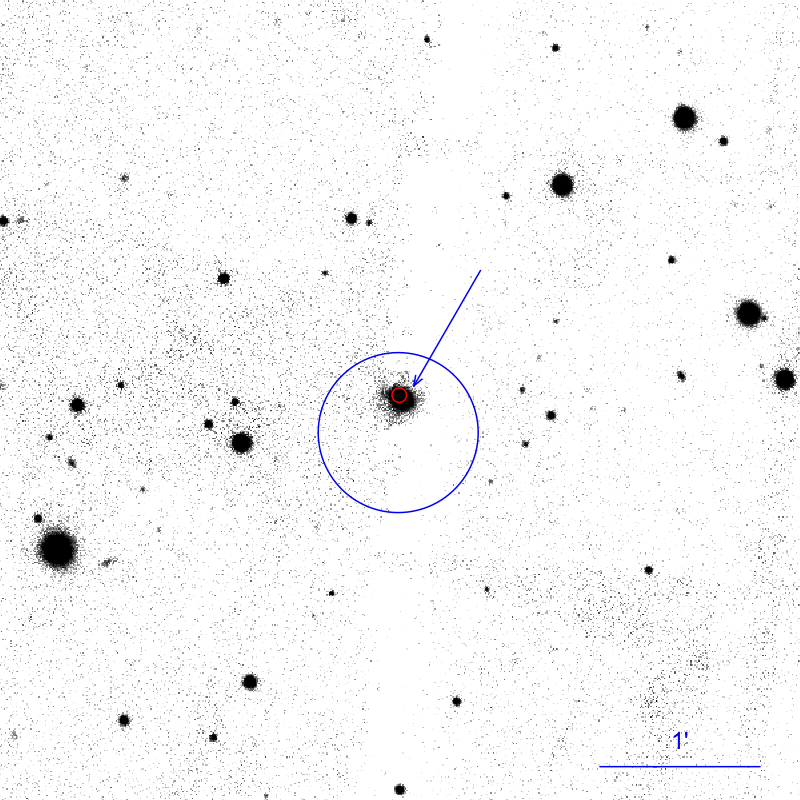}
   \includegraphics[width=0.4\columnwidth]{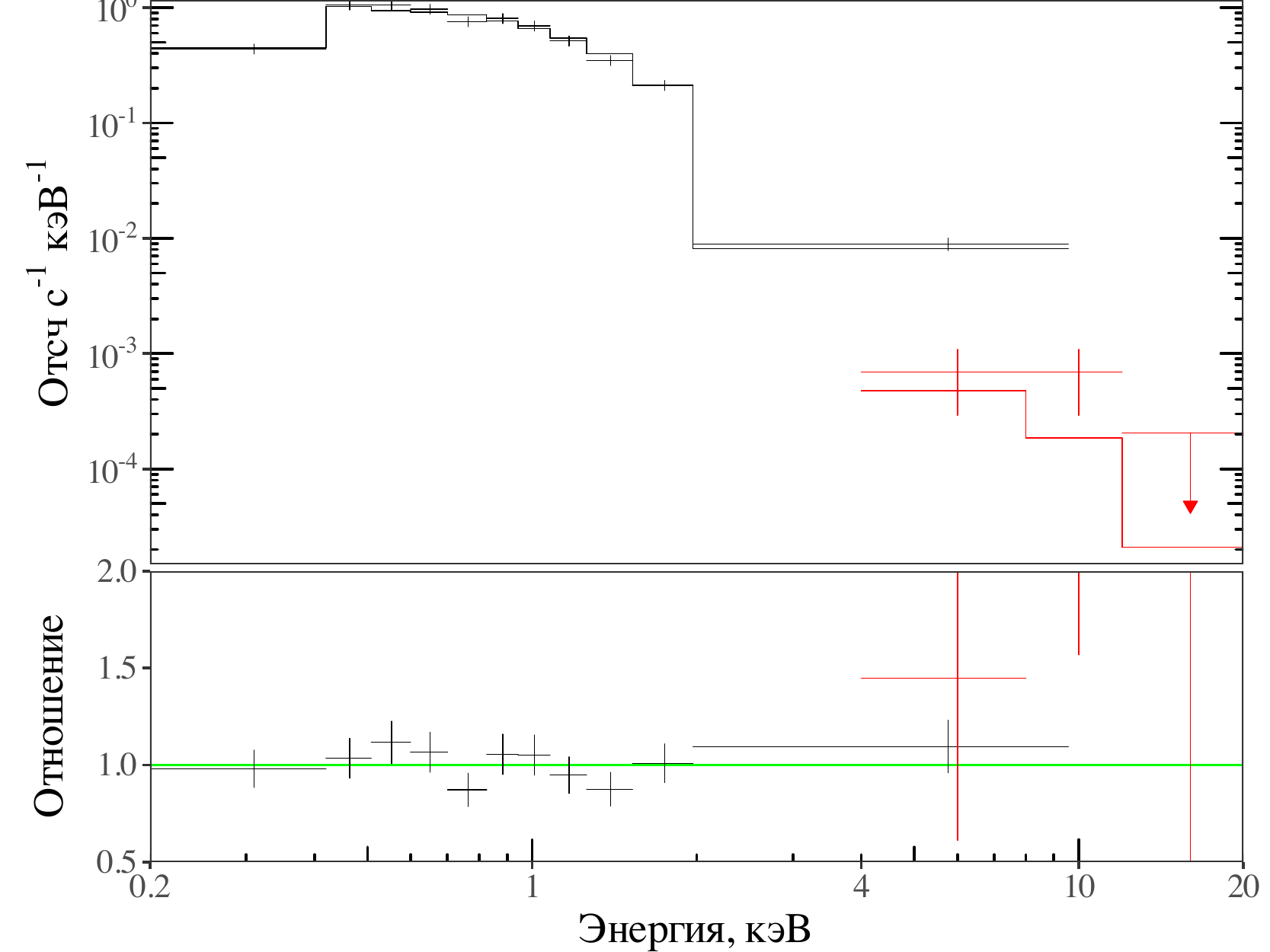}
  \end{floatrow}
  \vfill
  \vspace{1cm}
  \vfill
  \vspace{-1cm}
  \begin{floatrow}
  \includegraphics[width=0.6\columnwidth]{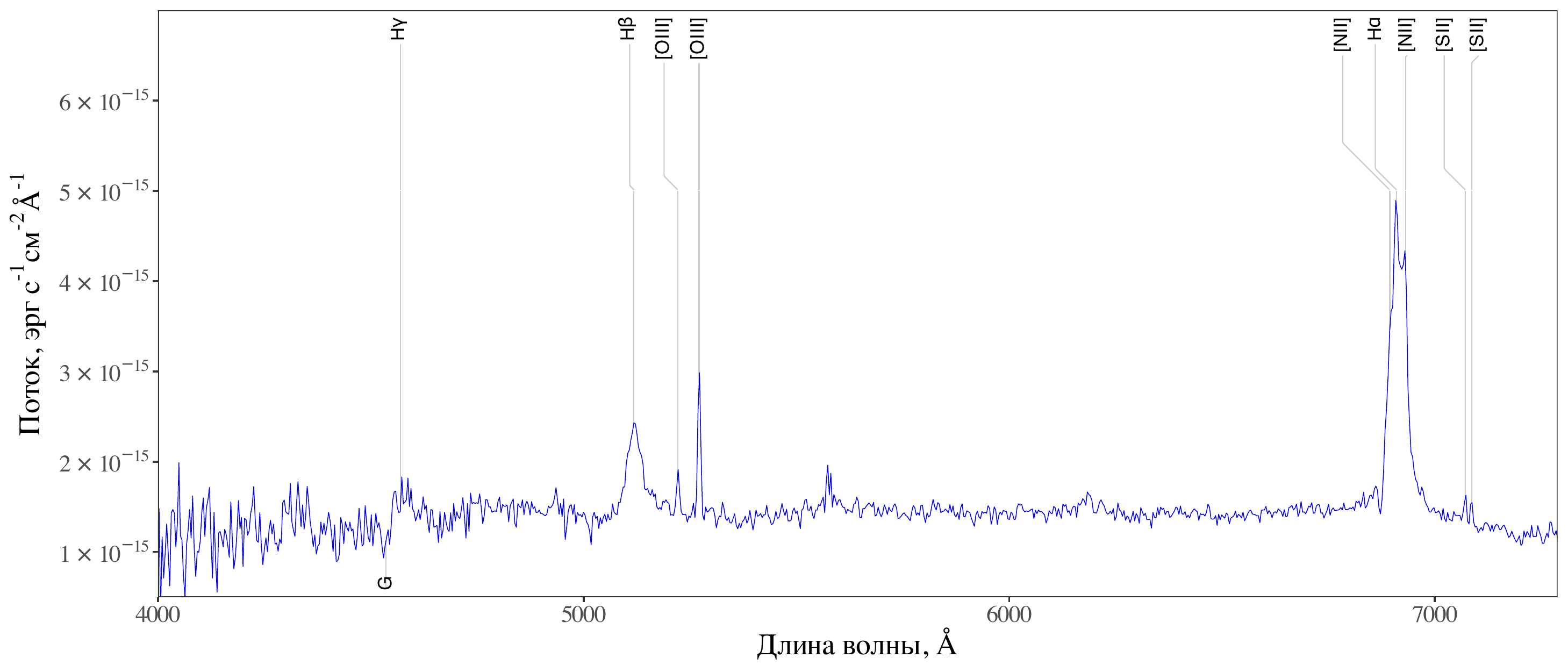}
  \end{floatrow}
  \caption{
  Same as Fig.~\ref{fig:spec0706}, but for SRGA\,J$092021.6\!+\!860249$.
  }
  \label{fig:spec0920}
\end{figure*}

This source was discovered during the \rosat\
all-sky survey (2RXS J092015.8+860253). There is the galaxy LEDA~2790304 with an infrared color typical for AGNs ($W1-W2=0.64$) in the \ero\ position error circle (Fig.~\ref{fig:spec0920}). The radio source NVSS\,J$091958\!+\!860300$ is also associated with it.

The optical spectrum (Fig.~\ref{fig:spec0920}, Table~\ref{tab:j0920}) exhibits the Balmer H${\alpha}$, H${\beta}$, and H${\gamma}$ emission lines with broad components, narrow forbidden [OIII]$\lambda$4959, 
[OIII]$\lambda$5007, [NII]$\lambda$6584, [SII]$\lambda$6718, [SII]$\lambda$6732 lines, and the G absorption line. The measured redshift of the object is $z = 0.05286\pm 0.00013$. The narrow-line flux ratios lg([OIII]$\lambda 5007/$H${\beta})=0.9\pm0.2$ and lg([NII]$\lambda 6584/$H${\alpha})>0.01$ are typical for AGNs, while the presence of broad H${\alpha}$, H${\beta}$ and H${\gamma}$ components allows the object to be classified as Sy1.

\begin{table*}
  \caption{Spectral features of SRGA\,J$092021.6\!+\!860249$} 
  \label{tab:j0920}
  \vskip 2mm
  \renewcommand{\arraystretch}{1.1}
  \renewcommand{\tabcolsep}{0.35cm}
  \centering
  \footnotesize
  \begin{tabular}{lcccc}
    \noalign{\doubleline}
    Line & Wavelength, \AA & Flux, $10^{-14}$~erg~s$^{-1}$~cm$^{-2}$ & Eq. width, \AA & $FWHM$, $10^2$ km/s\\
    \noalign{\vskip 3pt\hrule\vskip 5pt}
H$\gamma$, narrow      & 4581 & $<0.2$   & $>-1.6$ & $-$\\
H$\gamma$, broad    & 4581 & $1.8\pm0.4$   & $-14.3\pm 3.0$ & $32\pm7$\\
H$\beta$, narrow       & 5121 & $0.2\pm0.1$   & $- 1.1\pm 0.5$ & $-$\\
H$\beta$, broad     & 5121 & $4.2\pm0.2$   & $-30\pm 2$ & $28\pm2$\\
{}[OIII]$\lambda$4959 & 5222 & $0.4\pm0.1$   & $- 2.8\pm 0.5$ & $-$\\
{}[OIII]$\lambda$5007 & 5272 & $1.3\pm0.1$   & $- 9.0\pm 0.4$ & $-$\\
{}[NII]$\lambda$6548  & 6894 & $<0.7$        & $>-4.3$        & $-$                        \\
H$\alpha$, narrow      & 6910 & $<1.5$        & $>-9.9$        & $-$                        \\
H$\alpha$, broad    & 6910 & $10.5\pm1.6$  & $-70\pm11$ & $14.9\pm0.6$\\
{}[NII]$\lambda$6584  & 6931 & $2.1 \pm0.6$  & $-13.8\pm 3.8$ & $-$\\
{}[SII]$\lambda$6718  & 7072 & $0.19\pm0.04$ & $- 1.5\pm 0.3$ & $-$\\
{}[SII]$\lambda$6732  & 7086 & $0.16\pm0.03$ & $- 1.2\pm 0.3$ & $-$\\
    \noalign{\vskip 3pt\hrule\vskip 5pt}
  \end{tabular}
\end{table*}

\subsection{\it SRGAJ195702.4+615036}

\begin{figure*}
  \centering
  \vfill
  SRGA\,J$195702.4\!+\!615036$
  \vfill
  \vskip 0.5cm
  \begin{floatrow}
    \includegraphics[width=0.3\columnwidth]{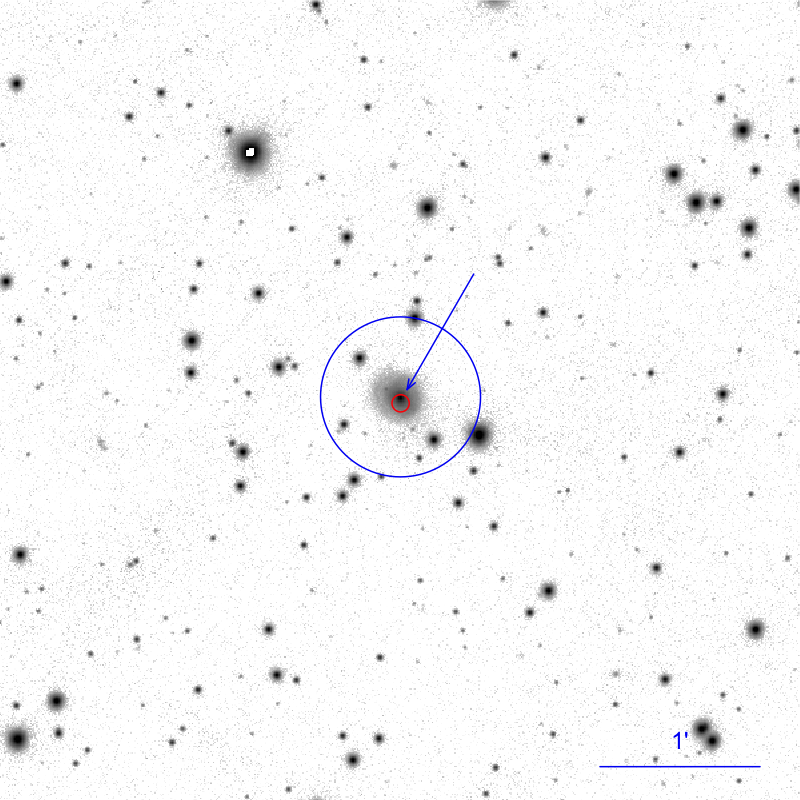}
   \includegraphics[width=0.4\columnwidth]{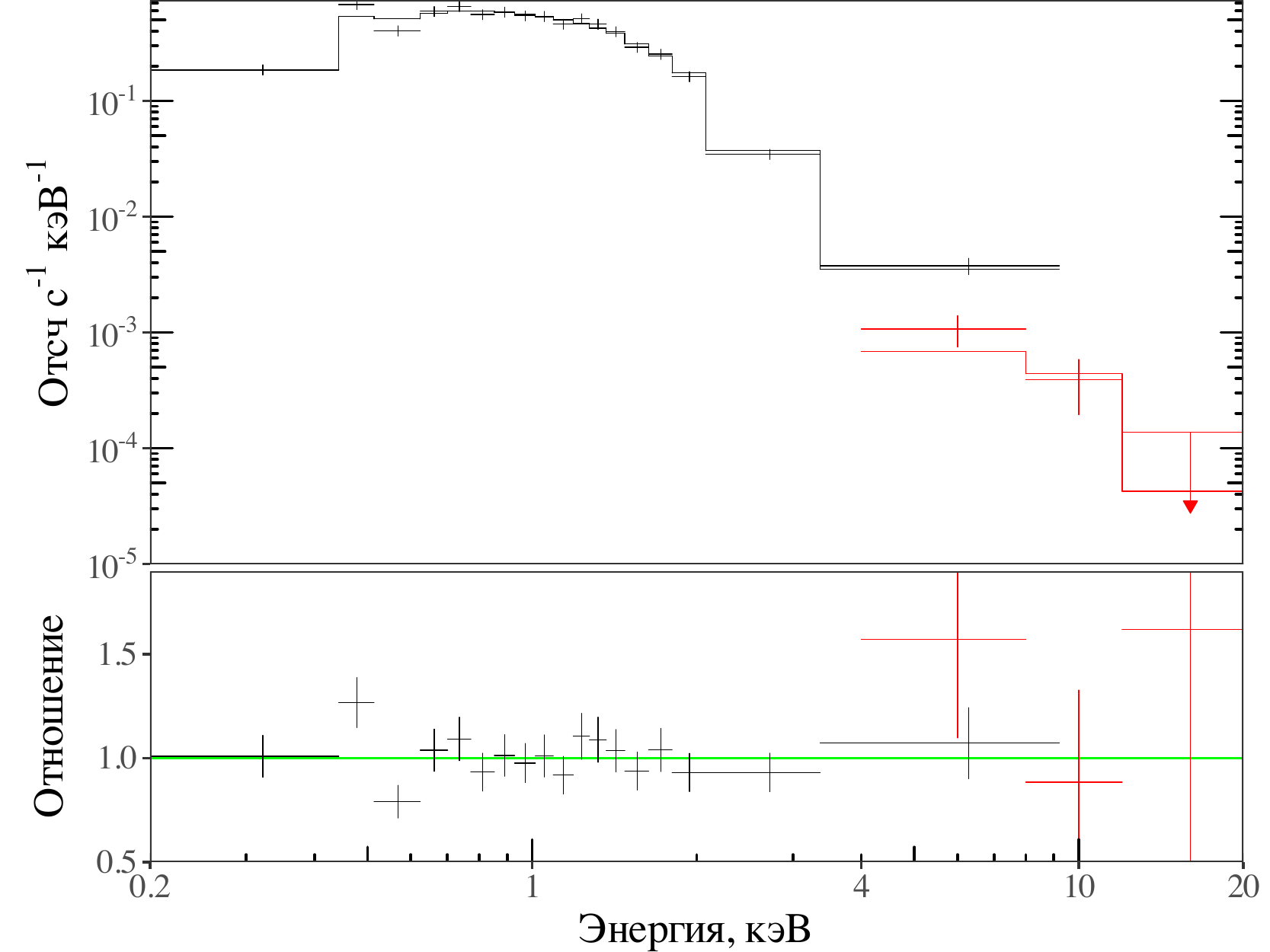}
  \end{floatrow}
  \vfill
  \vspace{1cm}
  \vfill
  \vspace{-1cm}
  \begin{floatrow}
  \includegraphics[width=0.6\columnwidth]{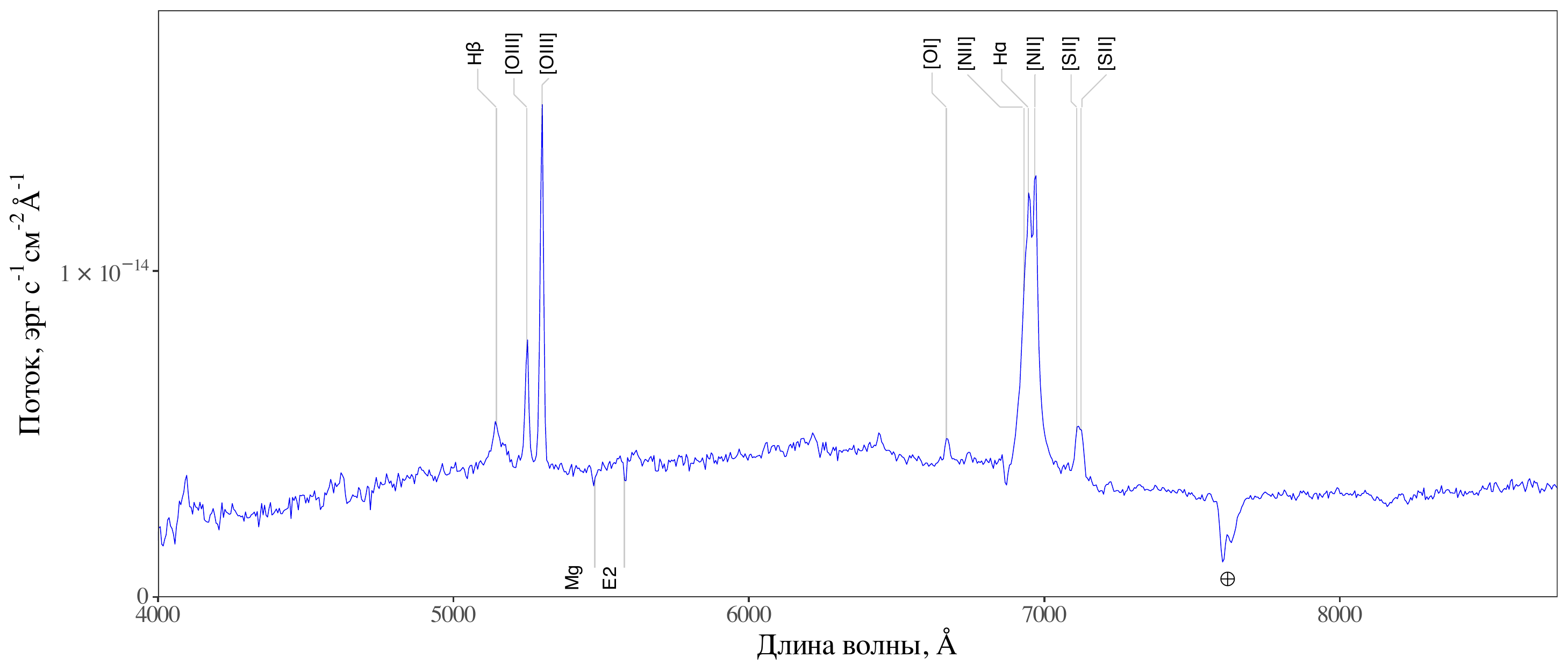}
  \end{floatrow}
  \caption{
  Same as Fig.~\ref{fig:spec0706}, but for SRGA\,J$195702.4\!+\!615036$.
  }
  \label{fig:spec1957}
\end{figure*}

This source was discovered during the ROSAT all-sky survey (2RXS J195702.4+615038). There is the galaxy LEDA~2625686 with an infrared color typical for AGNs ($W1-W2=0.65$) in the \ero\ position error circle (Fig.~\ref{fig:spec1957}).

The galaxy’s spectrum (Fig.~\ref{fig:spec1957}, Table~\ref{tab:j1957}) exhibits the H${\alpha}$ and H${\beta}$ emission lines with broad com- ponents, the forbidden [OIII]$\lambda$4959, [OIII]$\lambda$5007, 
[OI]$\lambda$6300, [SII]$\lambda$6718, [SII]$\lambda$6732 lines, and the Fraunhofer MgI absorption line. The measured redshift of the object is $z = 0.05857\pm0.00014$. The narrow-line flux ratios lg([OIII]$\lambda 5007/$H${\beta})=1.34\pm0.19$ and lg([NII]$\lambda 6584/$H${\alpha})=0.24\pm0.12$ point to the presence of an active galactic nucleus. The flux in the broad H${\alpha}$ and H${\beta}$ components exceeds considerably the flux in the narrow ones, which allows the object to be classified as Sy1.

\begin{table*}
  \caption{Spectral features of SRGA\,J$195702.4\!+\!615036$} 
  \label{tab:j1957}
  \vskip 2mm
  \renewcommand{\arraystretch}{1.1}
  \renewcommand{\tabcolsep}{0.35cm}
  \centering
  \footnotesize
  \begin{tabular}{lcccc}
    \noalign{\doubleline}
    Line & Wavelength, \AA & Flux, $10^{-14}$~erg~s$^{-1}$~cm$^{-2}$ & Eq. width, \AA & $FWHM$, $10^2$ km/s\\
    \noalign{\vskip 3pt\hrule\vskip 5pt}
H$\beta$, narrow       & 5146 & $0.7\pm0.3$  & $- 1.7\pm0.7$ & $-$\\
H$\beta$, broad     & 5146 & $5.0\pm0.7$  & $-12.5\pm1.7$ & $35\pm5$\\
{}[OIII]$\lambda$4959 & 5249 & $5.5\pm0.3$  & $-13.7\pm0.6$ & $-$\\
{}[OIII]$\lambda$5007 & 5299 & $14.5\pm0.2$ & $-36\pm1$ & $-$\\
{}[OI]$\lambda$6300   & 6671 & $1.2\pm0.1$  & $- 2.9\pm0.3$ & $-$\\
{}[NII]$\lambda$6548  & 6935 & $2.2\pm0.7$  & $- 5.5\pm1.7$ & $-$\\
H$\alpha$, narrow      & 6950 & $3.9\pm1.0$  & $-10.0\pm2.5$ & $-$\\
H$\alpha$, broad    & 6950 & $36\pm3$ & $-92\pm6$ & $27\pm2$\\
{}[NII]$\lambda$6584  & 6970 & $6.9\pm0.8$  & $-17.6\pm2.2$ & $-$\\
{}[SII]$\lambda$6718  & 7111 & $2.2\pm0.3$  & $- 5.8\pm0.7$ & $-$\\
{}[SII]$\lambda$6732  & 7126 & $2.0\pm0.3$  & $- 5.4\pm0.7$ & $-$\\
    \noalign{\vskip 3pt\hrule\vskip 5pt}
  \end{tabular}
\end{table*}

\subsection{\it SRGA J221913.2+362014}

This X-ray source was discovered by the \art\
telescope onboard the \srg observatory during the first year of its all-sky survey \citep{pavlinsky2022}. There is the optical–infrared–radio source NVSS\,J$221914\!+\!362011$~= WISEA\,J$221914.50\!+\!362010.5$ with an infrared color typical for AGNs ($W1-W2=1.2$) in the \art\ position error circle refined based on the \ero\ data (Fig.~\ref{fig:spec0706}).

\begin{figure*}
  \centering
  \vfill
  SRGA\,J$221913.2\!+\!362014$
  \vfill
  \vskip 0.5cm
  \begin{floatrow}
    \includegraphics[width=0.3\columnwidth]{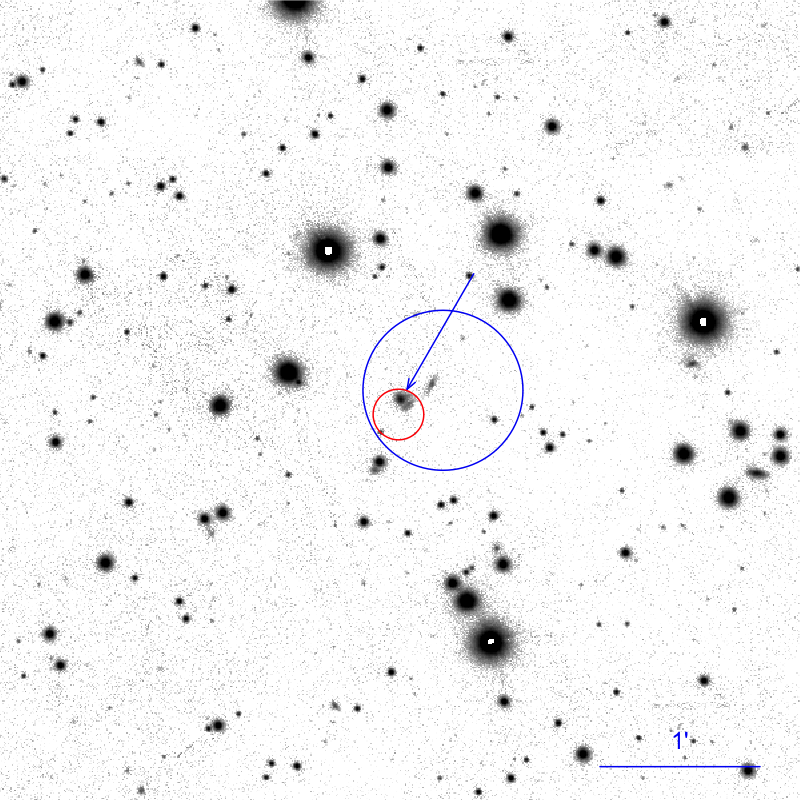}
   \includegraphics[width=0.4\columnwidth]{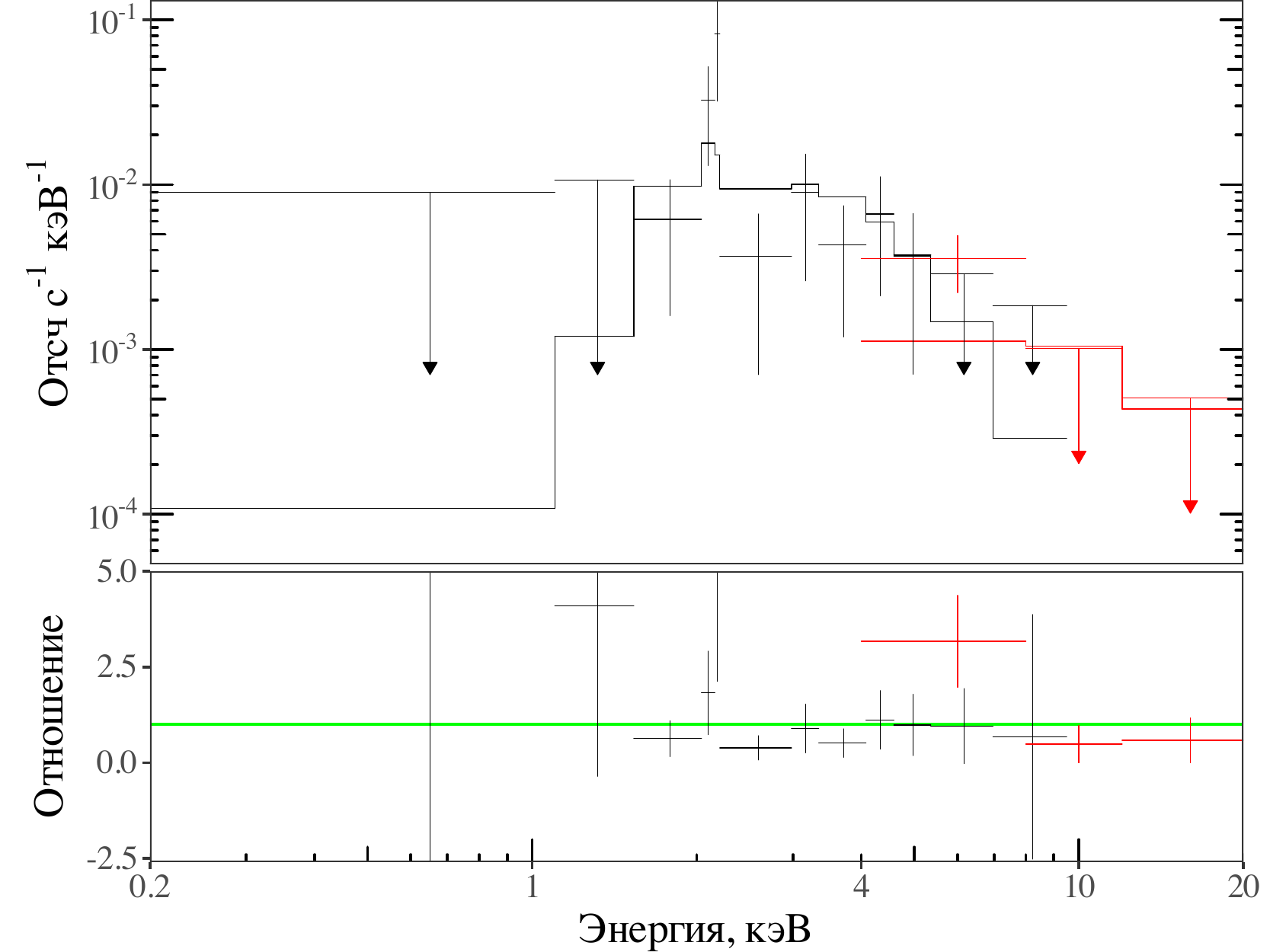}
   \includegraphics[width=0.37\columnwidth]{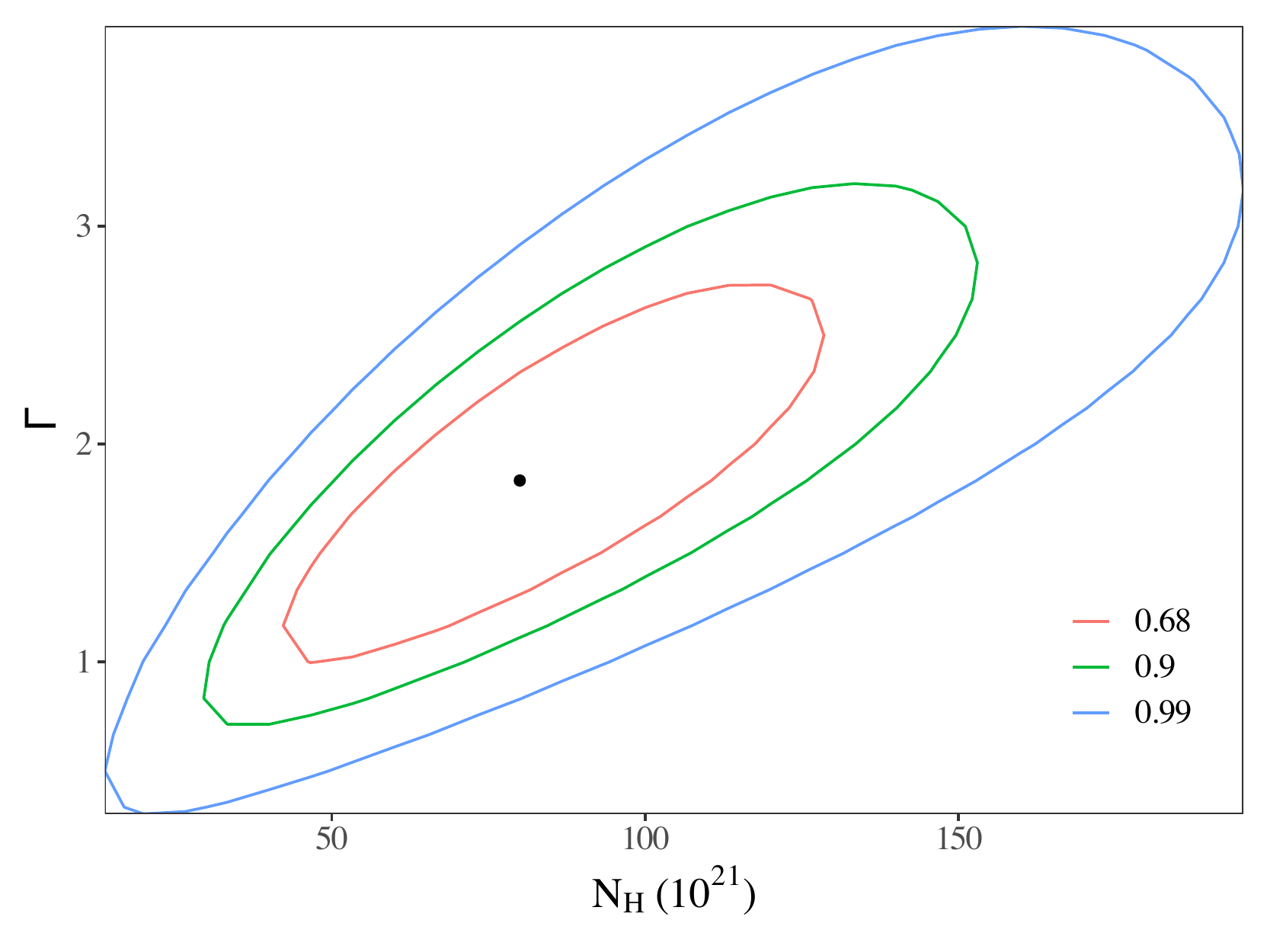}
  \end{floatrow}
  \vfill
  \vspace{1cm}
  \vfill
  \vspace{.1cm}
  \begin{floatrow}
  \includegraphics[width=0.35\columnwidth]{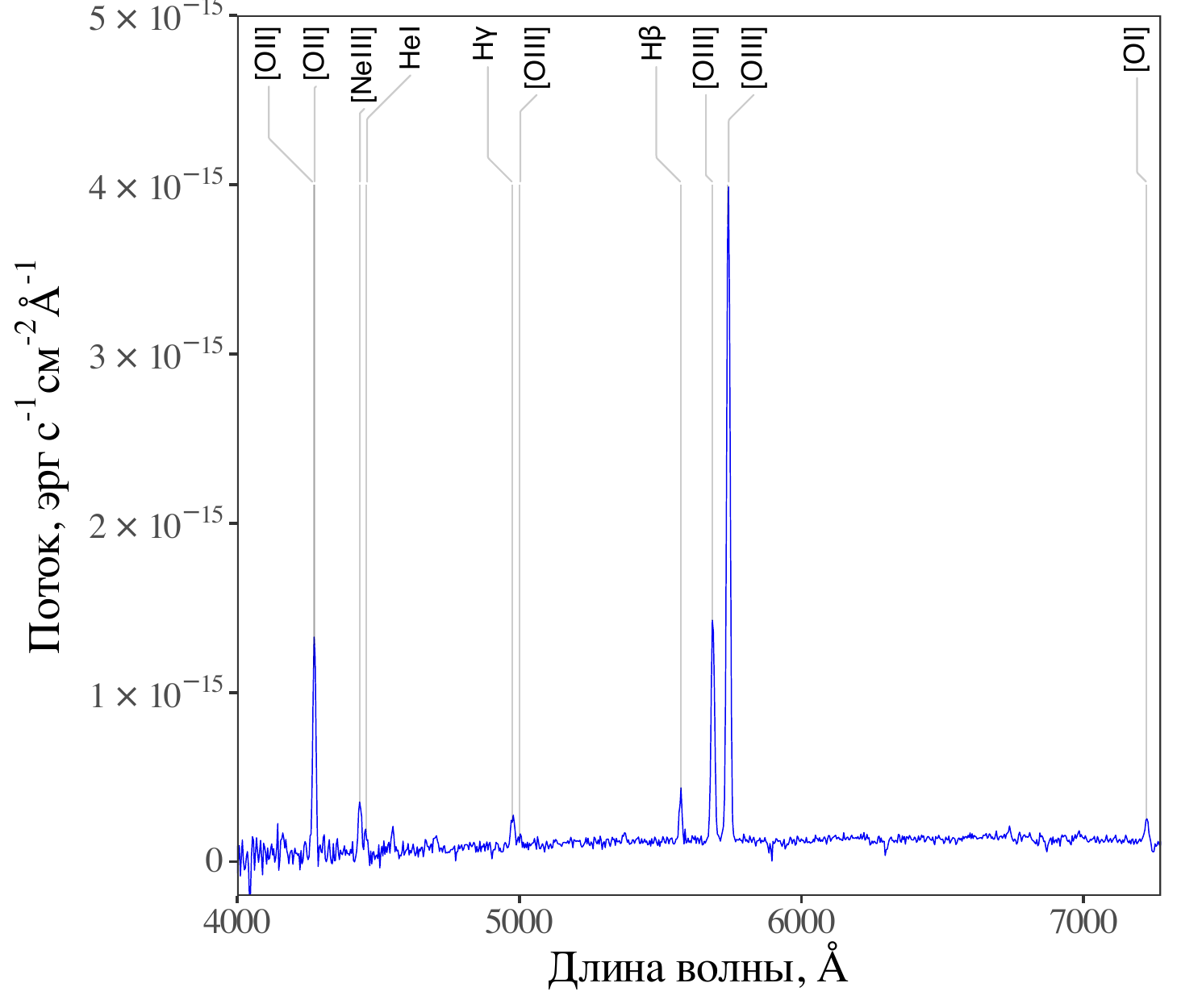}
  \includegraphics[width=0.35\columnwidth]{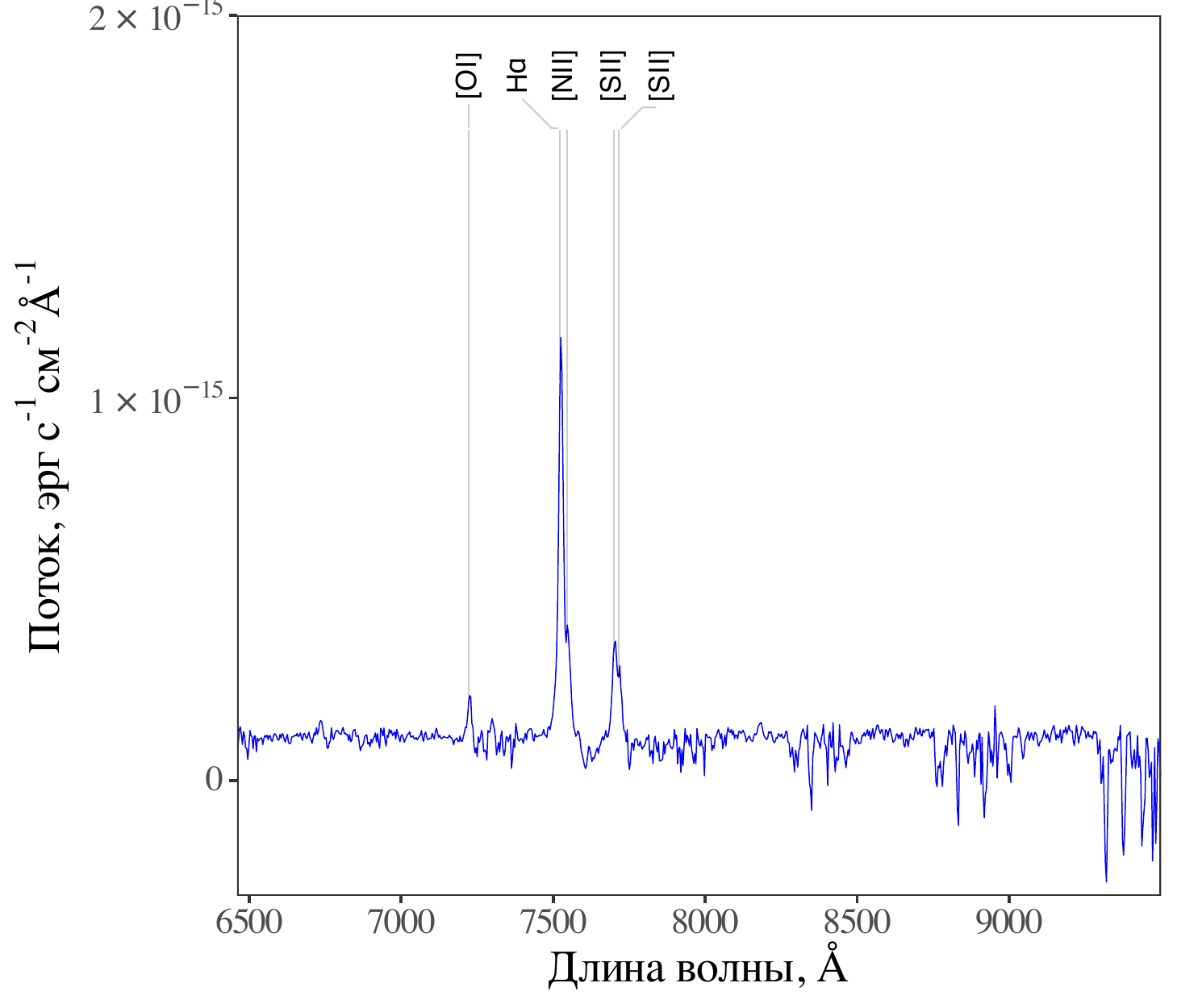}
  \end{floatrow}
  \caption{
  Same as Fig.~\ref{fig:spec0252}, but for SRGA\,J$221913.2\!+\!362014$. The red circumference in the pointing picture indicates the 98\% \ero\ source position error circle. The optical spectrum is shown on the two lower panels: the spectrum taken with VPGH600G (left) and the spectrum taken with VPHG600R (right)
  }
  \label{fig:spec2219}
\end{figure*}

The optical spectrum (Fig.~\ref{fig:spec2219}, Table~\ref{tab:j2219}) exhibits many forbidden neutral and ionized oxygen, sulfur, neon emission lines and narrow Balmer H${\alpha}$, H${\beta}$, and H${\gamma}$ emission lines. The measured redshift is $z = 0.14667\pm0.00003$. By its position on the BPT diagram (lg([OIII]$\lambda 5007/$H${\beta})=1.25\pm0.04$ and 
lg([NII]$\lambda 6584/$H${\alpha})=-0.58\pm0.04$), the object can be attributed to Seyfert galaxies, while the absence of broad Balmer line components implies that this is Sy2.

\begin{table*}
  \caption{Spectral features of SRGA\,J$221913.2\!+\!362014$} 
  \label{tab:j2219}
  \vskip 2mm
  \renewcommand{\arraystretch}{1.1}
  \renewcommand{\tabcolsep}{0.35cm}
  \centering
  \footnotesize
  \begin{tabular}{lcccc}
    \noalign{\doubleline}
    Line & Wavelength, \AA & Flux, $10^{-15}$~erg~s$^{-1}$~cm$^{-2}$ & Eq. width, \AA & $FWHM$, $10^2$ km/s\\
    \noalign{\vskip 3pt\hrule\vskip 5pt}
H$\beta$              & 5572 & $2.9\pm0.2 $ & $- 23\pm 2$ & $-$\\
{}[OIII]$\lambda$4959 & 5686 & $17.2\pm0.3$ & $-135\pm 2$ & $-$\\
{}[OIII]$\lambda$5007 & 5741 & $51\pm1$ & $-403\pm 2$ & $-$\\
{}[OI]$\lambda$6300   & 7225 & $1.5\pm0.2 $ & $- 13.8\pm 2.2$ & $-$\\
{}[NII]$\lambda$6548  & 7505 & $0.2\pm 0.1$ & $-12.4\pm 2.7$ & $-$\\
H$\alpha$             & 7526 & $18.0\pm0.8$ & $-132\pm 8$ & $-$\\
{}[NII]$\lambda$6584  & 7550 & $4.7\pm0.3 $ & $-35\pm 4$ & $-$\\
{}[SII]$\lambda$6718  & 7702 & $5.0\pm0.9 $ & $- 56\pm10$ & $-$\\
{}[SII]$\lambda$6732  & 7722 & $2.3\pm0.9 $ & $- 26\pm 10$ & $-$\\
    \noalign{\vskip 3pt\hrule\vskip 5pt}
  \end{tabular}
\end{table*}

\subsection{\it SRGA J223714.9+402939}

\begin{figure*}
  \centering
  \vfill
  SRGA\,J$223714.9\!+\!402939$
  \vfill
  \vskip 0.5cm
  \begin{floatrow}
    \includegraphics[width=0.3\columnwidth]{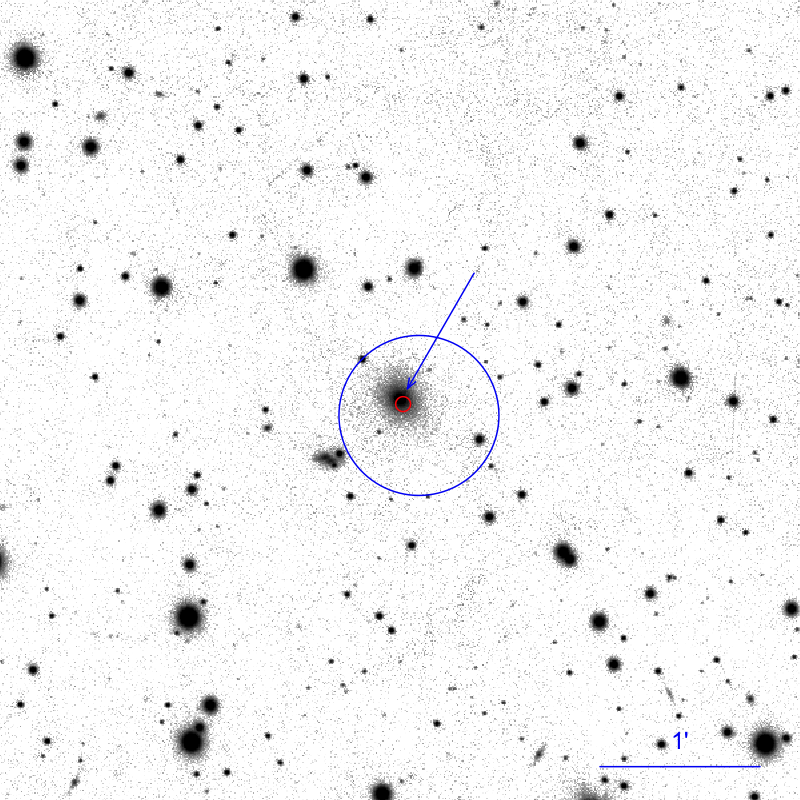}
   \includegraphics[width=0.4\columnwidth]{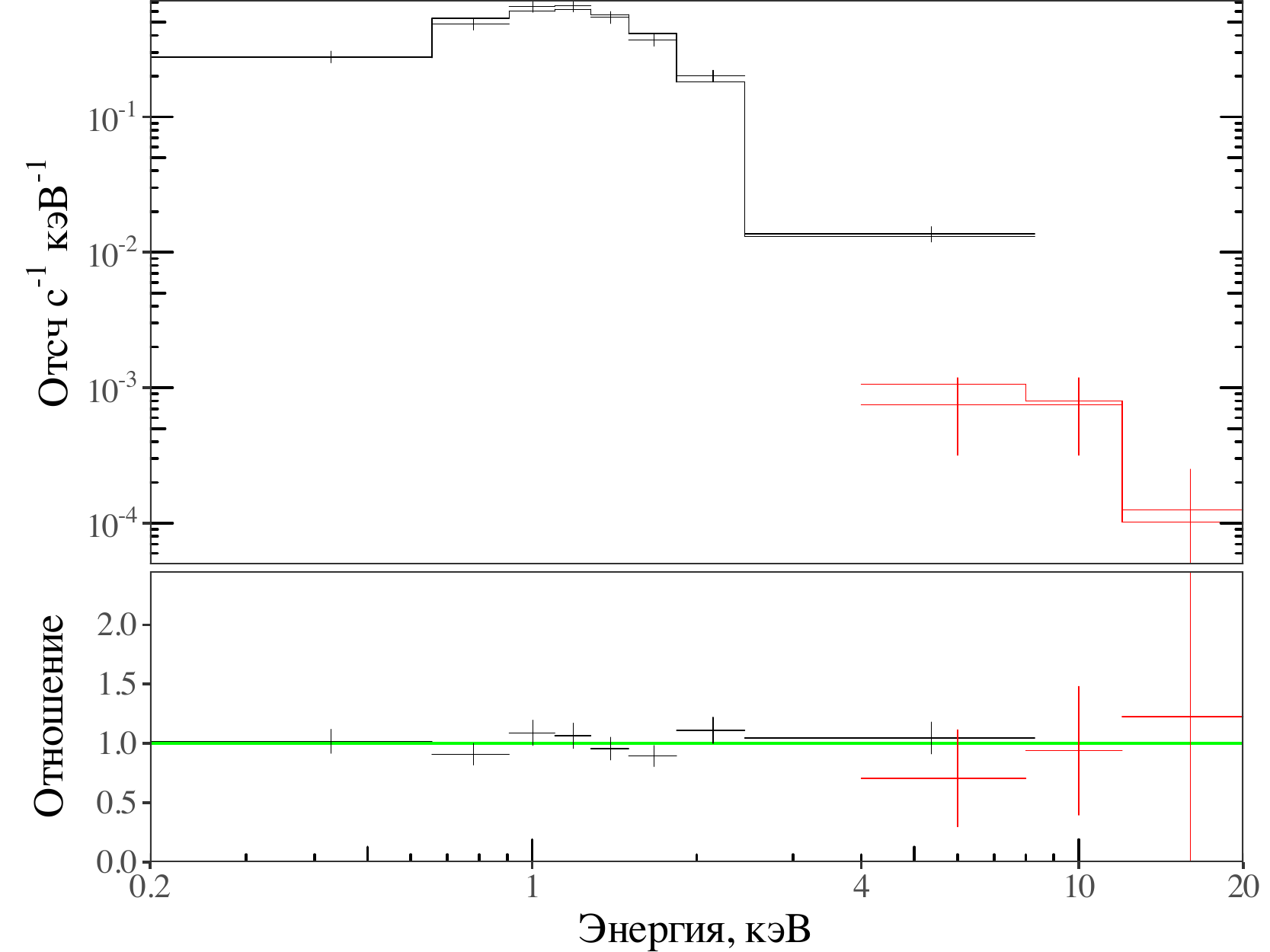}
  \end{floatrow}
  \vfill
  \vspace{1cm}
  \vfill
  \vspace{-1cm}
  \begin{floatrow}
  \includegraphics[width=0.6\columnwidth]{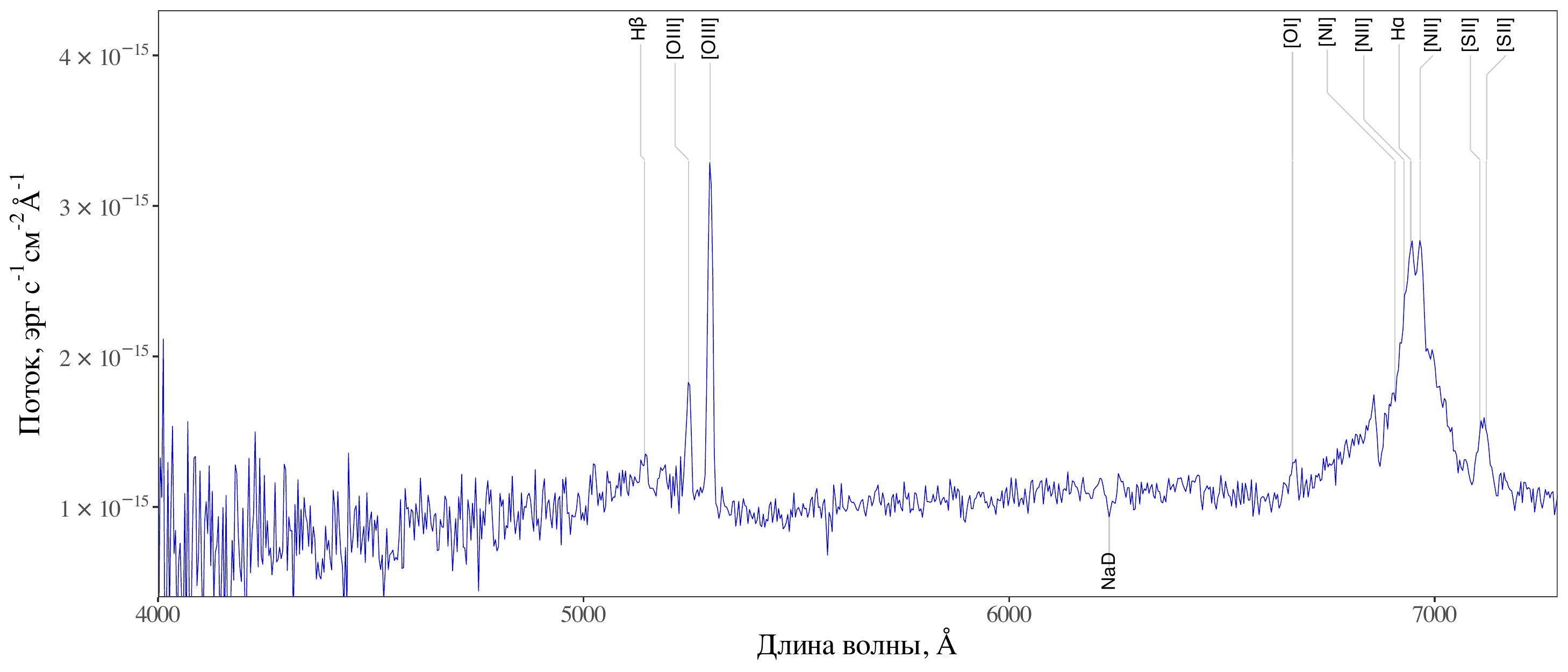}
  \end{floatrow}
  \caption{
   Same as Fig.~\ref{fig:spec0706}, but for SRGA\,J$223714.9\!+\!402939$.
  }
  \label{fig:spec2237}
\end{figure*}

This source was discovered during pointed ROSAT observations (1WGA J2237.2+4029) \citep{white2000}. There is the galaxy LEDA~5060459 with an infrared color typical for AGNs ($W1-W2=0.73$) in the \ero\ position error circle (Fig.~\ref{fig:spec2237}). According to SIMBAD, the galaxy’s redshift is $z=0.0580$.

The optical spectrum (Fig.~\ref{fig:spec2237}, Table~\ref{tab:j2237}) exhibits forbidden oxygen and sulfur emission lines as well as the H${\alpha}$ and H${\beta}$ emission lines with broad components. The refined redshift of the object is $z = 0.05818\pm0.00011$. By its position on the BPT diagram (lg([OIII]$\lambda 5007/$H${\beta})>1.14$ and
lg([NII]$\lambda 6584/$H${\alpha})=0.03\pm0.04$), the object can be attributed to Seyfert galaxies, while the presence of broad H${\alpha}$ and H${\beta}$ components, the flux in which is much greater than that in the narrow ones, allows the object to be classified as Sy1.

\begin{table*}
  \caption{Spectral features of SRGA\,J$223714.9\!+\!402939$} 
  \label{tab:j2237}
  \vskip 2mm
  \renewcommand{\arraystretch}{1.1}
  \renewcommand{\tabcolsep}{0.35cm}
  \centering
  \footnotesize
  \begin{tabular}{lcccc}
    \noalign{\doubleline}
    Line & Wavelength, \AA & Flux, $10^{-14}$~erg~s$^{-1}$~cm$^{-2}$ & Eq. width, \AA & $FWHM$, $10^2$ km/s\\
    \noalign{\vskip 3pt\hrule\vskip 5pt}
H$\beta$, narrow       & 5144 & $<0.2$       & $>-4.6$        & $-$ \\
H$\beta$, broad     & 5144 & $6.8\pm0.7$  & $- 73\pm8$ & $161\pm18$\\
{}[OIII]$\lambda$4959 & 5247 & $0.9\pm0.1$  & $-  9.5\pm1.3$ & $-$\\
{}[OIII]$\lambda$5007 & 5298 & $2.9\pm0.1$  & $- 31\pm2$ & $-$\\
{}[OI]$\lambda$6300   & 6668 & $0.30\pm0.07$& $-  2.8\pm0.7$ & $-$\\
{}[NII]$\lambda$6548  & 6929 & $0.6\pm0.1$  & $-  5.3\pm1.0$ & $*5.4$\\
H$\alpha$, narrow      & 6945 & $1.7\pm0.1$  & $- 15.5\pm1.1$ & $-$\\
H$\alpha$, broad    & 6945 & $17.5\pm1.0$ & $-164\pm9$ & $97\pm5$\\
{}[NII]$\lambda$6584  & 6967 & $1.8\pm0.1$  & $- 16.7\pm1.2$ & $-$\\
{}[SII]$\lambda$6718  & 7109 & $0.5\pm0.1$  & $-  4.5\pm0.6$ & $-$\\
{}[SII]$\lambda$6732  & 7124 & $0.5\pm0.1$  & $-  4.8\pm0.6$ & $-$\\
    \noalign{\vskip 3pt\hrule\vskip 5pt}
  \end{tabular}
  \begin{flushleft}
  * The value of the parameter was fixed when fitting the line.
  \end{flushleft}
\end{table*}

\subsection{\it SRGA J232037.8+482329}

This source was discovered during the ROSAT all-sky survey (2RXS J232039.7+482317). There is the galaxy LEDA~2316409 at $z=0.04150$ (according to SIMBAD) in the \ero\ position error circle, with which the radio source NVSS\,J$232039\!+\!482326$ is associated.

\begin{figure*}
  \centering
  \vfill
  SRGA\,J$232037.8\!+\!482329$
  \vfill
  \vskip 0.5cm
  \begin{floatrow}
    \includegraphics[width=0.3\columnwidth]{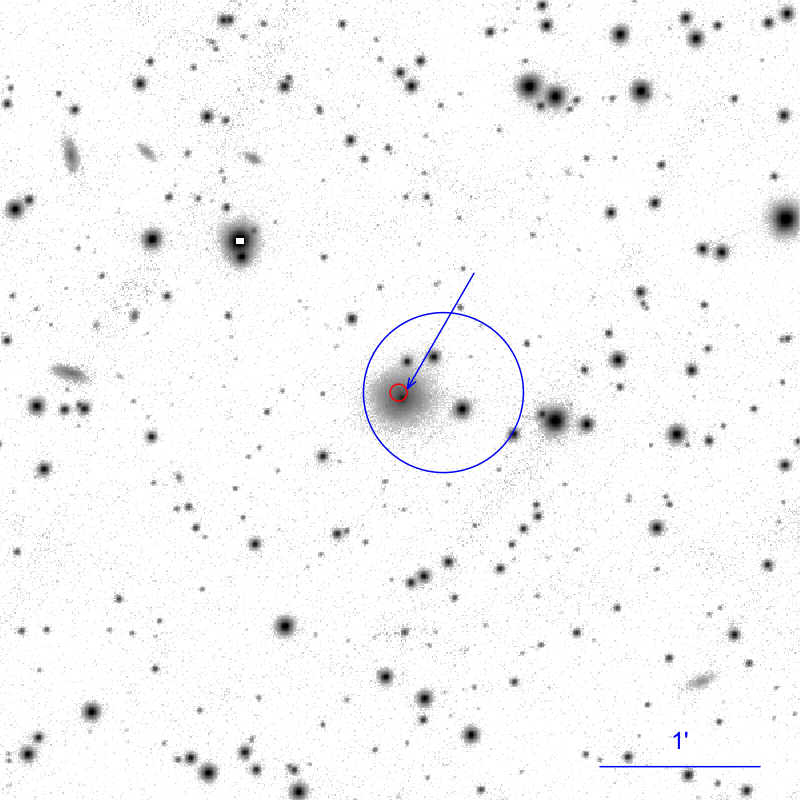}
   \includegraphics[width=0.4\columnwidth]{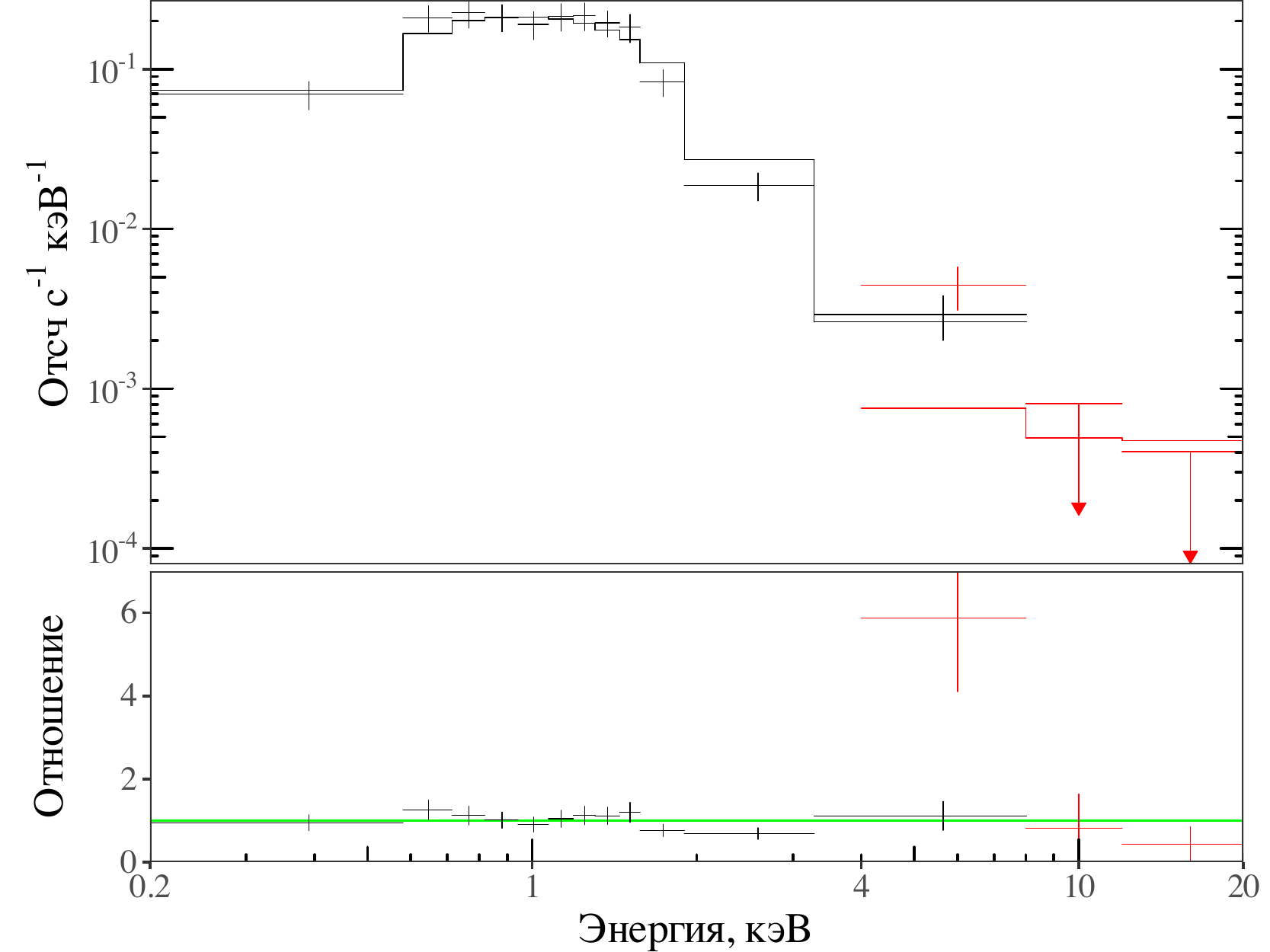}
  \end{floatrow}
  \vfill
  \vspace{1cm}
  \vfill
  \vspace{-1cm}
  \begin{floatrow}
  \includegraphics[width=0.6\columnwidth]{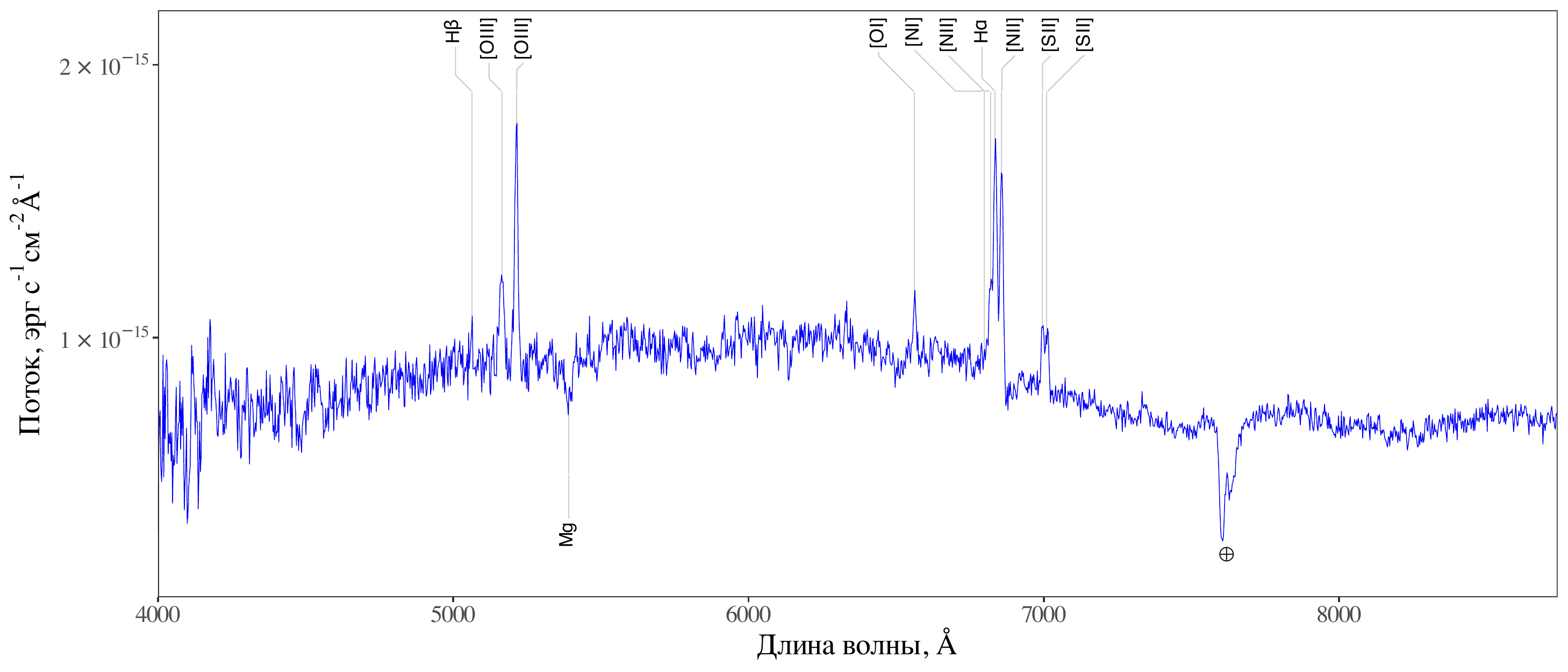}
  \end{floatrow}
  \caption{
  Same as Fig.~\ref{fig:spec0706}, but for SRGA\,J$232037.8\!+\!482329$.
  }
  \label{fig:spec2320}
\end{figure*}

The galaxy’s spectrum (Fig.~\ref{fig:spec2320}, Table~\ref{tab:j2320}) exhibits forbidden oxygen and sulfur lines, narrow H${\alpha}$ and H${\beta}$ emission lines, and the Fraunhofer MgI absorption line. The refined redshift of the galaxy is $z=0.04197\pm0.00017$. By its position on the BPT diagram (lg([OIII]$\lambda 5007/$H${\beta})=1.03\pm0.17$ and lg([NII]$\lambda6584/$H${\alpha})=-0.06\pm0.02$) and the absence of broad H${\alpha}$ and H${\beta}$ components, the object can be classified as Sy2.

\begin{table*}
  \caption{Spectral features of SRGA\,J$232037.8\!+\!482329$} 
  \label{tab:j2320}
  \vskip 2mm
  \renewcommand{\arraystretch}{1.1}
  \renewcommand{\tabcolsep}{0.35cm}
  \centering
  \footnotesize
  \begin{tabular}{lcccc}
    \noalign{\doubleline}
    Line & Wavelength, \AA & Flux, $10^{-15}$~erg~s$^{-1}$~cm$^{-2}$ & Eq. width, \AA & $FWHM$, $10^2$ km/s\\
    \noalign{\vskip 3pt\hrule\vskip 5pt}
H$\beta$              & 5062 & $1.1\pm0.4 $ & $- 1.2\pm0.5$ & $-$\\
{}[OIII]$\lambda$4959 & 5164 & $5.1\pm0.5 $ & $- 5.6\pm0.5$ & $-$\\
{}[OIII]$\lambda$5007 & 5213 & $11.5\pm0.5$ & $-12.7\pm0.6$ & $-$\\
{}[OI]$\lambda$6300   & 6563 & $2.9\pm0.4$ & $- 3.2\pm0.4$  & $-$\\
{}[NII]$\lambda$6548  & 6819 & $4.4\pm0.3$ & $- 5.0\pm0.4$  & $-$\\
H$\alpha$             & 6836 & $11.2\pm0.4$ & $-12.9\pm0.5$ & $-$\\
{}[NII]$\lambda$6584  & 6857 & $9.8\pm0.4$ & $-11.4\pm0.4$  & $-$\\
{}[SII]$\lambda$6718  & 6997 & $3.1\pm0.3$ & $- 3.8\pm0.4$  & $-$\\
{}[SII]$\lambda$6732  & 7012 & $2.8\pm0.3$ & $- 3.5\pm0.4$  & $-$\\
    \noalign{\vskip 3pt\hrule\vskip 5pt}
  \end{tabular}
\end{table*}

\subsection{\it SRGA J235250.6-170449}

\begin{figure*}
  \centering
  \vfill
  SRGA\,J$235250.6\!-\!170449$
  \vfill
  \vskip 0.5cm
  \begin{floatrow}
    \includegraphics[width=0.3\columnwidth]{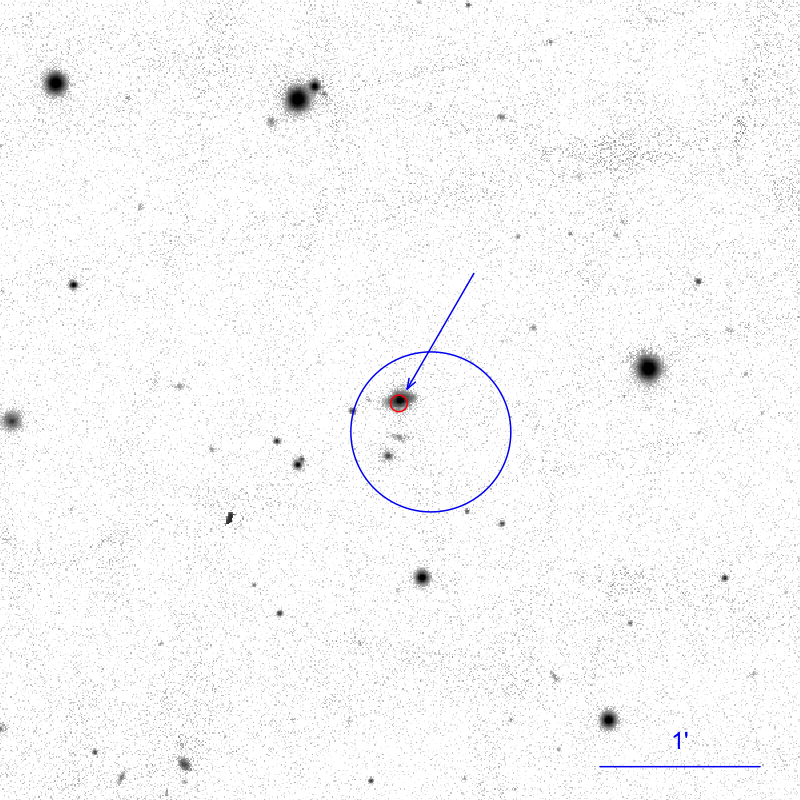}
   \includegraphics[width=0.4\columnwidth]{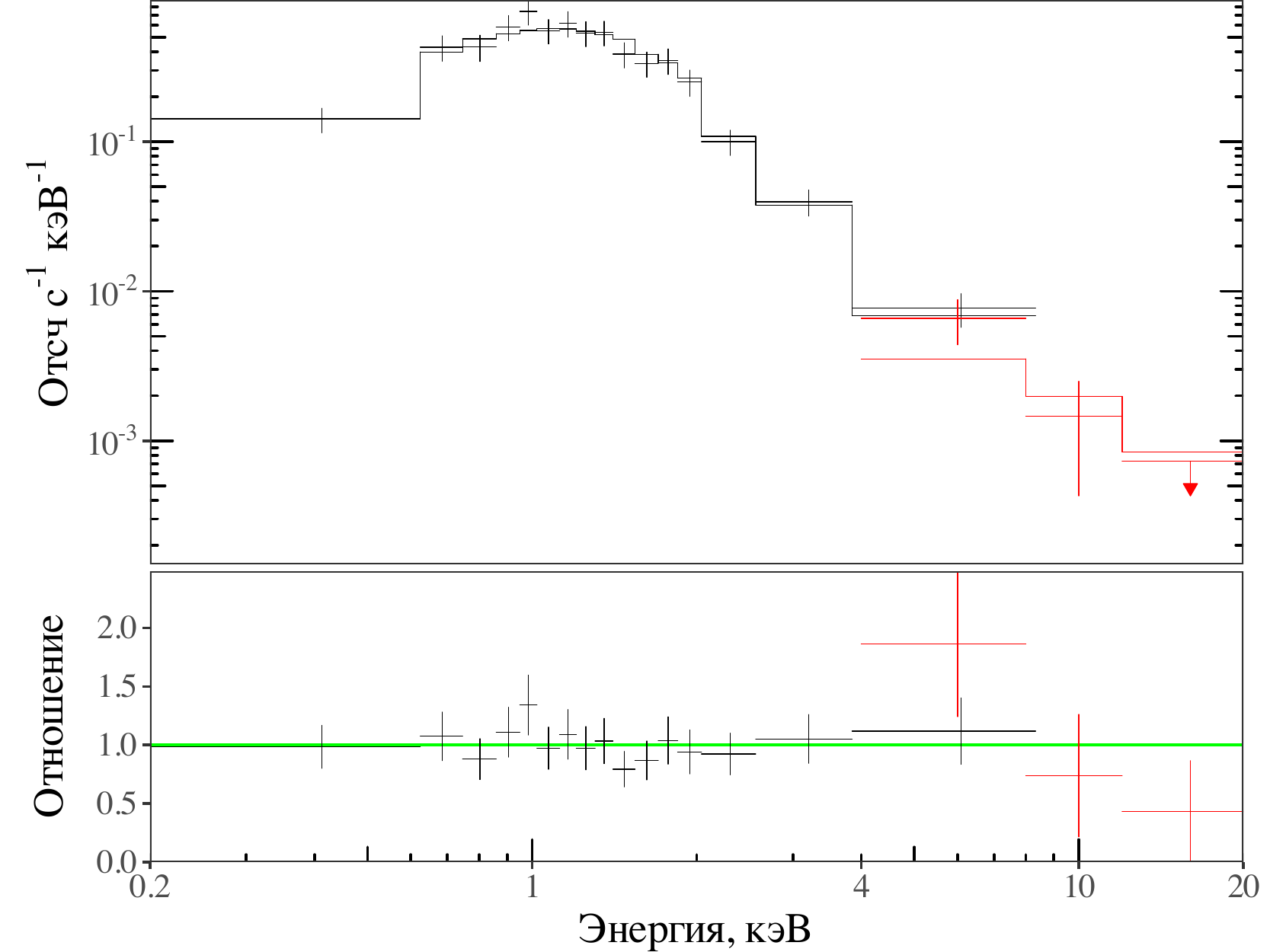}
   \includegraphics[width=0.35\columnwidth]{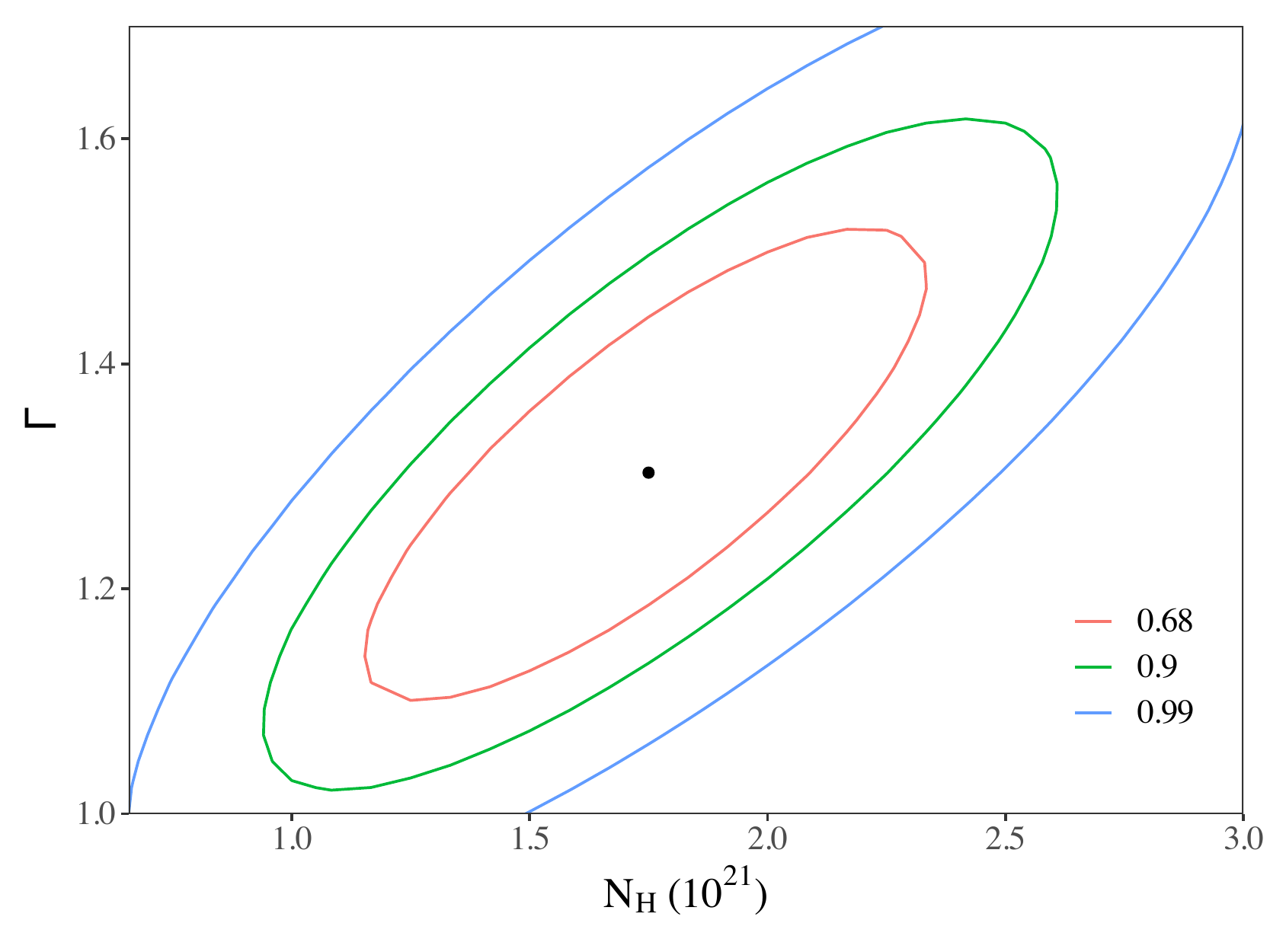}
  \end{floatrow}
  \vfill
  \vspace{1cm}
  \vfill
  \vspace{-1cm}
  \begin{floatrow}
  \includegraphics[width=0.6\columnwidth]{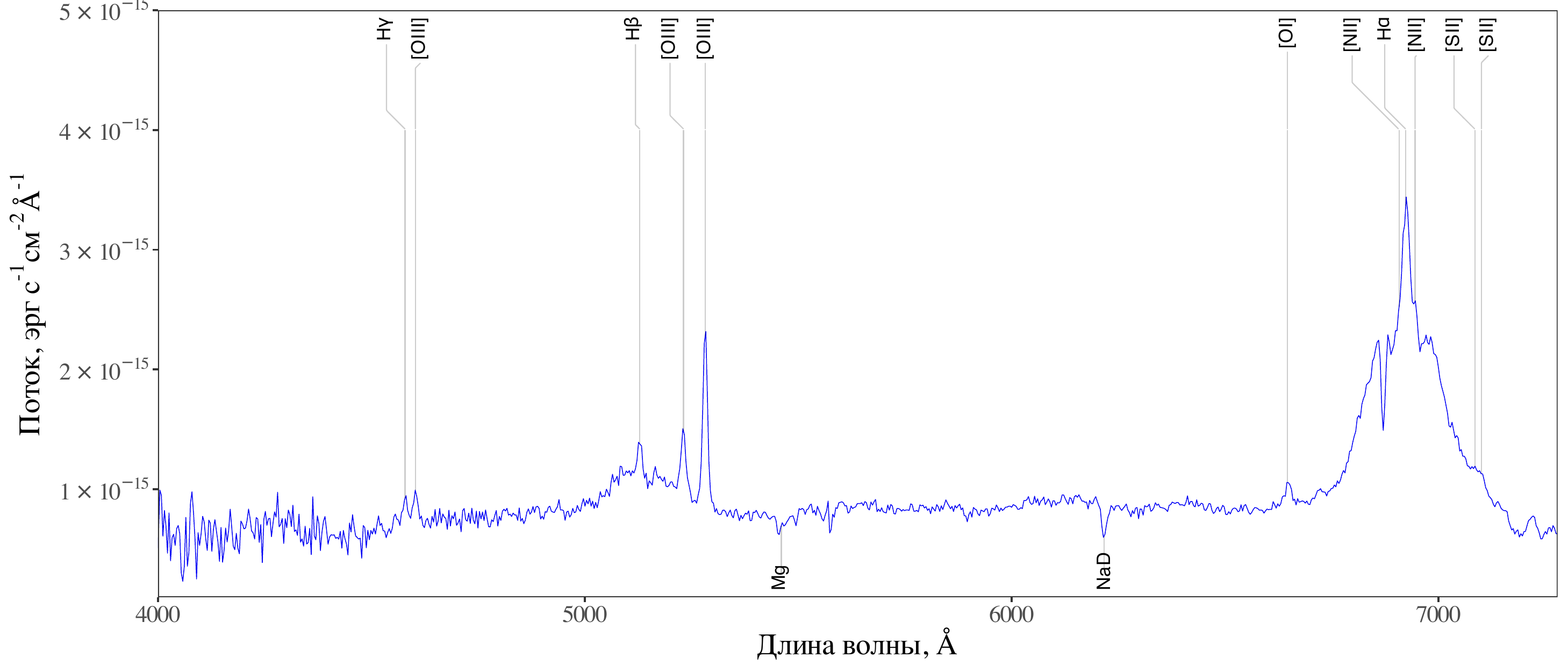}
  \end{floatrow}
  \caption{
  Same as Fig.~\ref{fig:spec0252}, but for SRGA\,J$235250.6\!-\!170449$. The red circumference in the pointing picture indicates the 98\% \ero\ source position error circle.
  }
  \label{fig:spec2352}
\end{figure*}

This source was discovered in the \swift/BAT hard X-ray sky survey \citep{oh2018}. There is the galaxy 2MASS\,J$23525142\!-\!1704372$ in the \ero\ position error circle (Fig.~\ref{fig:spec2352}), whose infrared color ($W1-W2=0.54$) suggests the presence of an AGN.

The galaxy’s spectrum (Fig.~\ref{fig:spec2352}, Table~\ref{tab:j2352}) exhibits many forbidden oxygen, sulfur, nitrogen emission lines and the Balmer H${\alpha}$, H${\beta}$, and H${\gamma}$ emission lines, the first two of which have intense broad components; the Fraunhofer MgI and NaD absorption lines are also seen. The measured red-shift of the object is $z=0.05502\pm 0.00012$. Although the object cannot be unambiguously attributed to Seyfert galaxies by its position on the BPT diagram (lg([OIII]$\lambda 5007/$H${\beta})=0.74\pm0.06$ and lg([NII]$\lambda 6584/$H${\alpha})=-1.07\pm0.22$), the presence of intense broad components in the H${\alpha}$ and H${\beta}$ emission lines allows it to be classified as Sy1.

\begin{table*}
  \caption{Spectral features of SRGA\,J$235250.6\!-\!170449$} 
  \label{tab:j2352}
  \vskip 2mm
  \renewcommand{\arraystretch}{1.1}
  \renewcommand{\tabcolsep}{0.35cm}
  \centering
  \footnotesize
  \begin{tabular}{lcccc}
    \noalign{\doubleline}
    Line & Wavelength, \AA & Flux, $10^{-15}$~erg~s$^{-1}$~cm$^{-2}$ & Eq. width, \AA & $FWHM$, $10^2$ km/s\\
    \noalign{\vskip 3pt\hrule\vskip 5pt}
H$\beta$, narrow       & 5129 & $3.5\pm0.5$  & $-  4.4\pm0.6$ & $-$\\
H$\beta$, broad     & 5129 & $67\pm3$ & $- 85\pm4$ & $112\pm5$\\
{}[OIII]$\lambda$4959 & 5231 & $7.6\pm0.5$  & $-  9.6\pm0.7$ & $-$\\
{}[OIII]$\lambda$5007 & 5282 & $19.0\pm0.5$ & $- 24\pm1$ & $-$\\
{}[OI]$\lambda$6302   & 6648 & $3.9\pm0.6$  & $-  4.7\pm0.7$ & $-$\\
{}[NII]$\lambda$6548  & 6909 & $1.1\pm0.7$  & $-  1.4\pm1.0$ & $-$\\
H$\alpha$, narrow      & 6925 & $15.2\pm0.8$ & $- 20\pm1$ & $-$\\
H$\alpha$, broad    & 6925 & $368\pm4$& $-493\pm6$ & $93\pm2$\\
{}[NII]$\lambda$6584  & 6946 & $1.3\pm0.6$  & $-  1.7\pm0.9$ & $-$\\
{}[SII]$\lambda$6718  & 7089 & $2.0\pm0.7$  & $-  2.8\pm0.9$ & $-$\\
{}[SII]$\lambda$6732  & 7105 & $2.5\pm0.6$  & $-  3.6\pm0.9$ & $-$\\
    \noalign{\vskip 3pt\hrule\vskip 5pt}
  \end{tabular}
\end{table*}

\subsection{\it Southern-Sky Objects}

\begin{figure*}
  \centering
  \vskip 0.5cm
  \includegraphics[width=0.32\columnwidth]{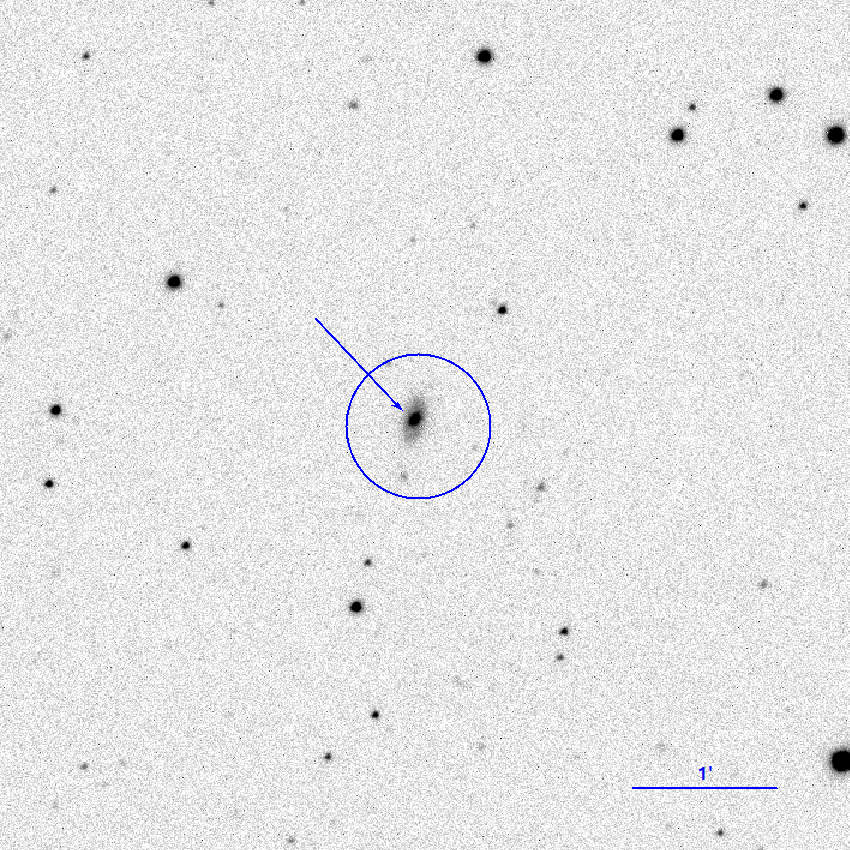}
  \includegraphics[width=0.32\columnwidth]{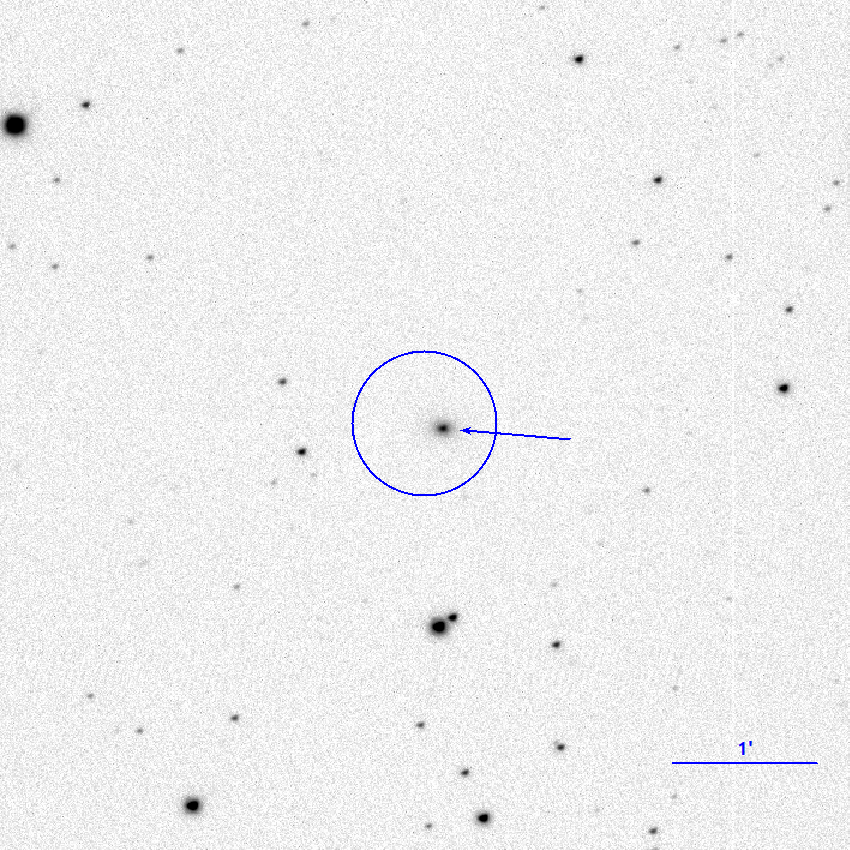}
  \includegraphics[width=0.32\columnwidth]{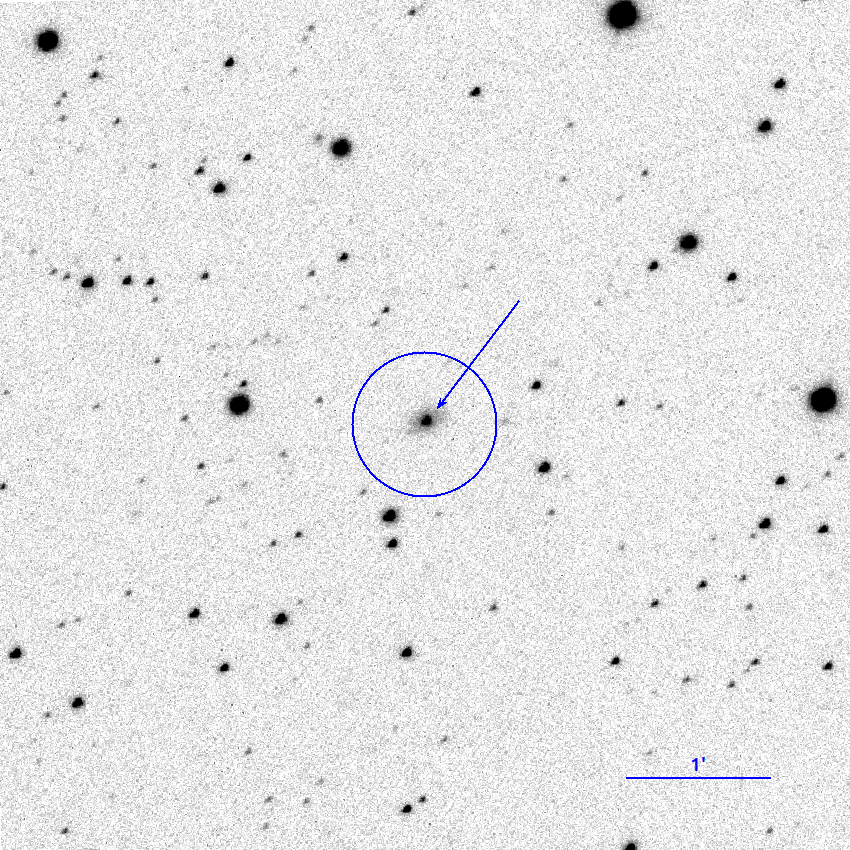}
  \vfill
  \vspace{0.1cm}
  SRGAJ030838.1-552041 \hspace{1.5cm} SRGAJ052959.8-340157 \hspace{1.5cm} SRGAJ055053.7-621457
  \vspace{0.7cm}
  \vfill
  \includegraphics[width=0.32\columnwidth]{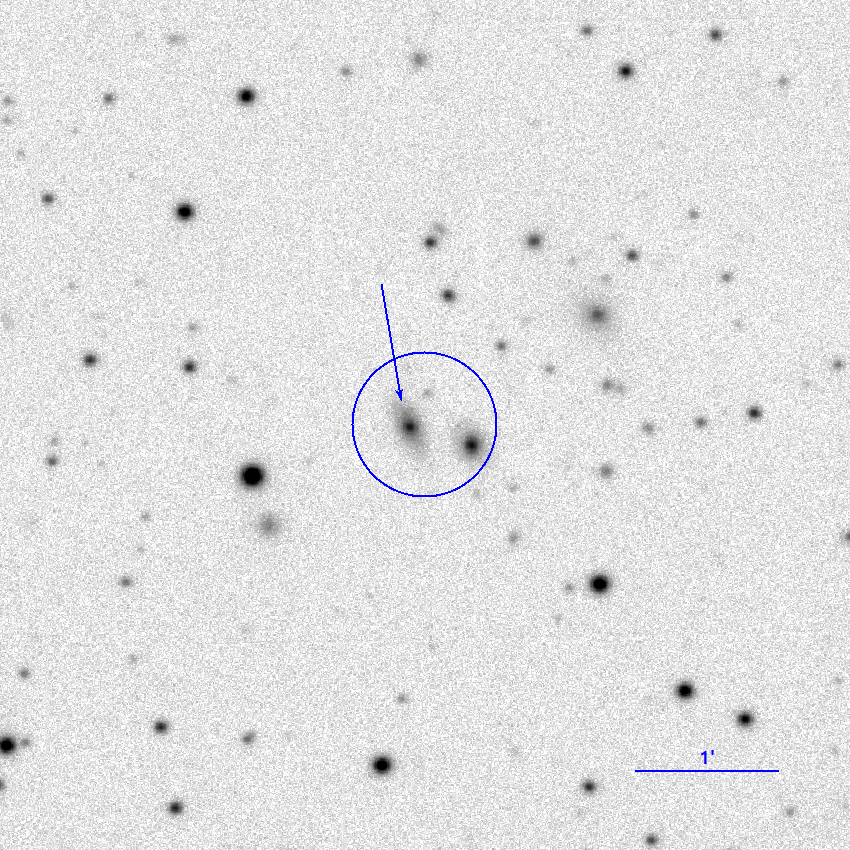}
  \includegraphics[width=0.32\columnwidth]{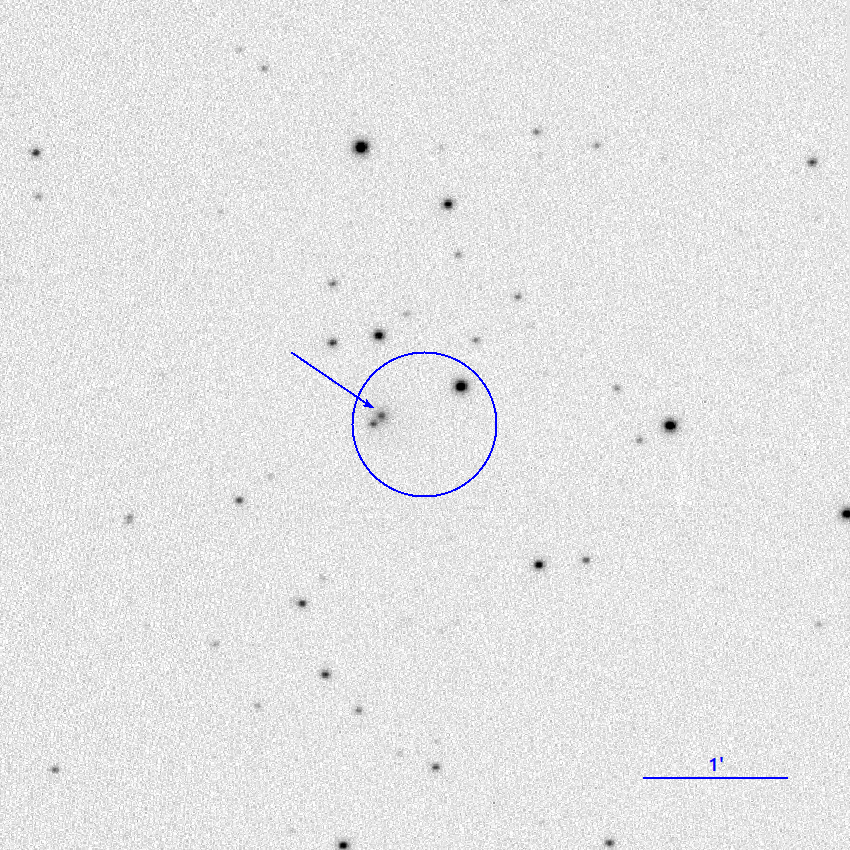}
  \includegraphics[width=0.32\columnwidth]{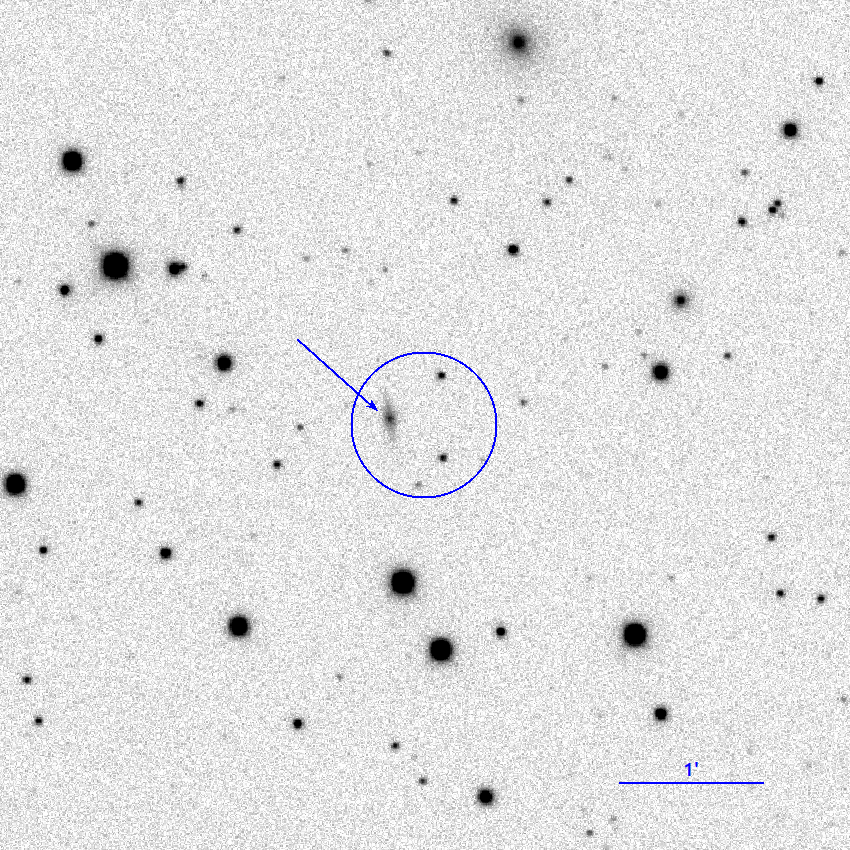}
  \vfill
  \vspace{0.1cm}
  SRGAJ060241.1-595152 \hspace{1.5cm} SRGAJ061322.9-290027 \hspace{1.5cm} SRGAJ063324.9-561424
  \vspace{0.7cm}
  \vfill
  \includegraphics[width=0.32\columnwidth]{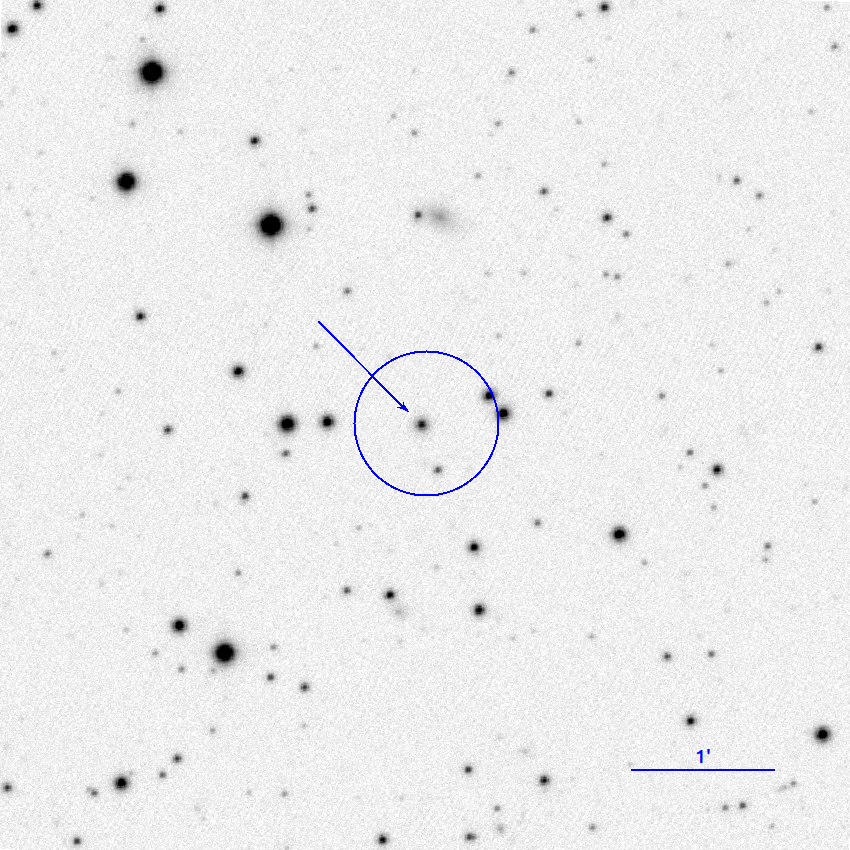}
  \includegraphics[width=0.32\columnwidth]{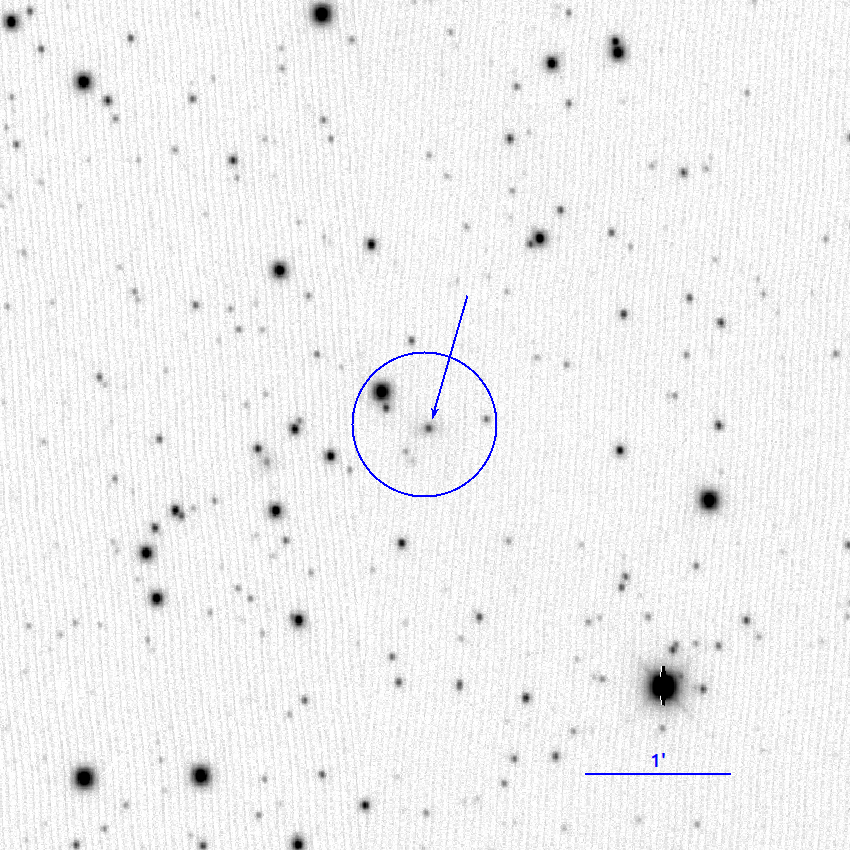}
  \vfill
  \vspace{0.1cm}
  SRGAJ064421.5-662620 \hspace{1.5cm} SRGAJ072823.5-440823
  \vspace{0.7cm}
  \vfill
  \caption{
  Optical images in the \emph{i} filter from the \emph{SkyMapper} survey \citep{keller2007} around eight \art\ X-ray sources in the southern sky. The blue circumference indicates the \art\ position error circle, with a radius of $30\arcsec$. The arrow indicates the galaxies for which a spectrum was taken in the \emph{6dF} survey.
  }
  \label{fig:chart_6df}
\end{figure*}

The eight \art\ southern-sky ($\delta>-20^\circ$) objects are identified with galaxies (Fig.~\ref{fig:chart_6df}), for which there are the spectra taken during the 6dF survey \citep{jones2004, jones2009}. One of these X-ray sources (SRGA J030838.1-552041) associated with the galaxy LEDA~410289 was discovered by the \art\ telescope onboard the \srg observatory \citep{pavlinsky2022}, two (SRGAJ052959.8–340157 = XMMSL2 J052958.9–340159 = LEDA 668116 and SRGAJ060241.1–595152 = XMMSL2 J060241.6– 595149 = LEDA 178859) were detected for the first time during the \xmm\ slew survey \citealt{xmm1018}, and the remaining five (SRGAJ055053.7-621457 = 2RXS J055054.2-621454 = LEDA 178653, 
SRGAJ061322.9-290027 = 2RXS J061324.1-290029 = LEDA 734640, 
SRGAJ063324.9-561424 = 2RXS J063326.4-561427 = LEDA 148903, 
SRGAJ064421.5-662620 = 2RXS J064422.7-662623 = 2MASS J06442187-6626199, 
SRGAJ072823.5-440823 = 2RXS J072822.3-440821 = 2MASS J07282338–4408241) were discovered during the ROSAT all-sky survey \citep{boller2016}. All of the listed galaxies are characterized by an infrared color typical for Seyfert galaxies ($W_1-W2$ from 0.5 to 1.0).

The 6dF survey was conducted at the UKST 1.2-m Schmidt telescope using a multi-fiber spectrograph with a $5.7^{\circ}$ field of view equipped with two low-resolution ($R\approx 1000$) gratings with overlapping spectral ranges. The range $4000$--$7500$\,\AA was completely covered. The spectra taken during the survey were not flux-calibrated and are presented\footnote{http://www-wfau.roe.ac.uk/6dFGS/} in counts,
which does not allow the absolute fluxes in emission lines to be measured. However, these data can be used to estimate the line equivalent widths and the ratios of the fluxes in the pairs of closely spaced lines (H${\beta}$, [OIII]$\lambda5007$) 
and ([NII]$\lambda6584$, H${\alpha}$), which are used to classify the AGNs by the BPT diagram.

Tables~\ref{tab:spec:6df}, \ref{tab:spec2:6df} and \ref{tab:bpt:6df} present the emission-line characteristics for the objects from our sample determined based on the 6dF spectra. The errors are given at the 68\% confidence level. The line FWHMs were corrected for the instrumental broadening of 5.75\,\AA. The redshifts of the objects were taken from the 6dF catalog \citep{jones2009}.

\begin{table}
  \caption{Spectral features of the objects from the 6dF survey} 
  \label{tab:spec:6df}
  \vskip 2mm
  \renewcommand{\arraystretch}{1.0}
  \renewcommand{\tabcolsep}{0.08cm}
  \centering
  \footnotesize
  \begin{tabular}{lcc}
    \noalign{\doubleline}
    Line & Eq. width, \AA & $FWHM$, $10^2$~km/s\\
    \noalign{\vskip 3pt\hrule\vskip 5pt}
\multicolumn{3}{c}{SRGA\,J$030838.1\!-\!552041$}\\
H${\beta}$ & $-3.2\pm 1.3$ & $2.8\pm1.1$\\
{}[OIII]$\lambda$4959 & $-12.6\pm 4.0$ & $3.5\pm1.1$\\
{}[OIII]$\lambda$5007 & $-41\pm 7$ & $3.5\pm0.3$\\
{}[NII]$\lambda$6548 & $-4.3^{+3.0}_{-2.4}$ & $3.5\pm1.3$\\
H${\alpha}$ & $-18.6\pm 2.0$ & $4.2\pm0.4$\\
{}[NII]$\lambda$6584 & $-14.3\pm 3.2$ & $3.7\pm0.5$\\
{}[SII]$\lambda$6717 & $-5.0\pm 1.6$ & $4.1\pm1.7$\\
{}[SII]$\lambda$6730 & $-4.6\pm 1.6$ & $3.9\pm1.7$\\
\multicolumn{3}{c}{SRGA\,J$052959.8\!-\!340157$}\\
{}[OII]$\lambda$3727 & $-16.6\pm 4.9$ & $5.8\pm0.9$\\
H${\beta}$ & $-4.6\pm 1.8$ & $6.0\pm1.7$\\
{}[OIII]$\lambda$4959 & $-21\pm 4$ & $6.9\pm0.6$\\
{}[OIII]$\lambda$5007 & $-65\pm 9$ & $6.3\pm0.6$\\
{}[NII]$\lambda$6548 & $- 4.1\pm0.9$ & $6.3\pm0.3$\\
H${\alpha}$, narrow & $-20.6\pm0.9$ & $6.3\pm0.3$\\
H${\alpha}$, broad & $-62\pm5$	& $72\pm6$ \\
{}[NII]$\lambda$6584 & $-16.2\pm0.9$ & $6.3\pm0.3$\\
\multicolumn{3}{c}{SRGA\,J$055053.7\!-\!621457$}\\
H${\delta}$, broad & $-17.5\pm1.9$ & $32\pm4$\\
H${\gamma}$, broad & $-16.2\pm1.2$ & $22\pm2$\\
H${\beta}$, narrow & $<1.9$ & -\\
H${\beta}$, broad & $-38\pm2$ & $24\pm2$ \\
{}[OIII]$\lambda$4959 & $-6.3\pm 2.2$ & $6.7\pm1.6$\\
{}[OIII]$\lambda$5007 & $-21\pm 2$ & $6.7\pm0.6$\\
H${\alpha}$, narrow & $- 6.4\pm2.8$ & $5.6\pm0.8$\\
H${\alpha}$, broad & $-90\pm5$ & $20\pm1$\\
{}[NII]$\lambda$6584 & $-12.8\pm2.4$ & $5.5\pm0.8$\\
\multicolumn{3}{c}{SRGA\,J$060241.1\!-\!595152$}\\
H${\beta}$ & $-6.7\pm 2.3$ & $3.1\pm0.8$\\
{}[OIII]$\lambda$4959 & $-14.0\pm 3.7$ & $4.4\pm0.5$\\
{}[OIII]$\lambda$5007 & $-49\pm 11$ & $3.7\pm0.4$\\
H${\alpha}$ & $-19^{+9}_{-18}$ & $4.9\pm1.7$\\
{}[NII]$\lambda$6584 & $-14^{+7}_{-14}$ & $4.9\pm1.7$\\
    \noalign{\vskip 3pt\hrule\vskip 5pt}
  \end{tabular}
\end{table}

\begin{table}
  \caption{ Continuation} 
  \label{tab:spec2:6df}
  \vskip 2mm
  \renewcommand{\arraystretch}{1.1}
  \renewcommand{\tabcolsep}{0.08cm}
  \centering
  \footnotesize
  \begin{tabular}{lcc}
    \noalign{\doubleline}
    Line & Eq. width, \AA & $FWHM$, $10^2$~km/s\\
    \noalign{\vskip 3pt\hrule\vskip 5pt}
\multicolumn{3}{c}{SRGA\,J$061322.9\!-\!290027$}\\
H${\beta}$ & $-3.1\pm 1.2$ & $4.1\pm1.5$\\
{}[OIII]$\lambda$4959 & $-8.2\pm 2.8$ & $5.6\pm1.3$\\
{}[OIII]$\lambda$5007 & $-25\pm 5$ & $5.1\pm0.5$\\
{}[OI]$\lambda$6300 & $-5.2\pm 1.9$ & $6.3\pm1.3$\\
{}[NII]$\lambda$6548 & $-7.9\pm 2.4$ & $6.1\pm0.8$\\
H${\alpha}$ & $-16.8\pm 2.4$ & $6.1\pm0.8$\\
{}[NII]$\lambda$6584 & $-21\pm 3$ & $6.0\pm0.8$\\
{}[SII]$\lambda$6717 & $-5.4^{+1.6}_{-3.0}$ & $5.4\pm2.1$\\
{}[SII]$\lambda$6730 & $-8.2\pm 3.1$ & $7.6\pm2.0$\\
\multicolumn{3}{c}{SRGA\,J$063324.9\!-\!561424$}\\
{}[OIII]$\lambda$4959 & $-7.8^{+3.6}_{-5.4}$ & $3.0\pm1.2$\\
{}[OIII]$\lambda$5007 & $-29^{+10}_{-27}$ & $3.0\pm1.2$\\
{}[NII]$\lambda$6548 & $-11.8\pm 3.8$ & $4.2\pm1.4$\\
H${\alpha}$ & $-23^{+6}_{-16}$ & $4.2\pm1.4$\\
{}[NII]$\lambda$6584 & $-15.3\pm 4.9$ & $4.2\pm1.4$\\
\multicolumn{3}{c}{SRGA\,J$064421.5\!-\!662620$}\\
H${\gamma}$ & $-2.9\pm 0.9$ & $3.7\pm1.7$\\
{}[OII]$\lambda$3727 & $-14.4\pm 4.8$ & $7.6\pm1.6$\\
H${\beta}$, narrow & $-5.2\pm 1.4$ & $2.9\pm1.6$\\
H${\beta}$, broad & $-35\pm2$ & $58\pm4$\\
{}[OIII]$\lambda$4959 & $-8.6\pm 1.8$ & $4.0\pm0.9$\\
{}[OIII]$\lambda$5007 & $-31\pm 6$ & $4.7\pm0.8$\\
H${\alpha}$, narrow & $-22\pm1$ & $2.8\pm0.2$\\
H${\alpha}$, broad & $-130\pm4$ & $40\pm2$\\
{}[NII]$\lambda$6584 & $-12.2\pm0.8$  & $2.8\pm0.2$ \\
\multicolumn{3}{c}{SRGA\,J$072823.5\!-\!440823$}\\
H${\gamma}$, broad & $-24\pm2$ & $32\pm3$\\
H${\beta}$, narrow & $<1.4$ & -\\
H${\beta}$, broad & $-55\pm2$ & $34\pm2$\\
{}[OIII]$\lambda$4959 & $-4.7^{+1.1}_{-3.4}$ & $4.6\pm1.6$\\
{}[OIII]$\lambda$5007 & $-20^{+4}_{-6}$ & $4.3\pm0.9$\\
H${\alpha}$, narrow & $-12.5\pm2.7$ & $5.0\pm0.8$\\
H${\alpha}$, broad & $-210\pm5$ & $30\pm1$\\
{}[NII]$\lambda$6584 & $- 5.4\pm1.8$ & $5.0\pm0.8$\\
    \noalign{\vskip 3pt\hrule\vskip 5pt}
  \end{tabular}
\end{table}

\begin{table*}
  \caption{Emission-line flux ratios for the objects from the 6dF survey} 
  \label{tab:bpt:6df}
  \vskip 2mm
  \renewcommand{\arraystretch}{1.1}
  \renewcommand{\tabcolsep}{0.35cm}
  \centering
  \footnotesize
  \begin{tabular}{ccc}
    \noalign{\doubleline}
    Object & lg([OIII]$\lambda5007/$H${\beta})$ & lg([NII]$\lambda6584/$H${\alpha})$\\
    \noalign{\vskip 3pt\hrule\vskip 5pt}
  SRGA\,J$030838.1\!-\!552041$ & $1.05\pm0.12$ & $-0.11\pm0.07$\\
  SRGA\,J$052959.8\!-\!340157$ & $1.09\pm0.07$ & $-0.11\pm0.03$\\
  SRGA\,J$055053.7\!-\!621457$ & $>1.11$ & $0.30\pm0.21$\\
  SRGA\,J$060241.1\!-\!595152$ & $0.89\pm0.13$ & $-0.15\pm0.37$\\
  SRGA\,J$061322.9\!-\!290027$ & $0.96\pm0.15$ & $0.09\pm0.08$\\
  SRGA\,J$063324.9\!-\!561424$ & $>0.87$ & $-0.19\pm0.24$\\
  SRGA\,J$064421.5\!-\!662620$ & $0.85\pm0.04$ & $-0.26\pm0.03$\\
  SRGA\,J$072823.5\!-\!440823$ & $>1.14$ & $-0.37\pm0.17$\\
    
    \noalign{\vskip 3pt\hrule\vskip 5pt}
  \end{tabular}
\end{table*}

\section{PROPERTIES OF THE DETECTED AGNs}

Tables~\ref{tab:list2} and \ref{tab:list2_6df} present basic properties of the AGNs that we managed to identify in this paper: the redshift, the optical type, and the X-ray luminosity $\lx$ in the 4--12~keV energy band.

We found the X-ray luminosity based on the 4--12~keV flux (see Tables~\ref{tab:list_src} and \ref{tab:list_src_6df}) from the catalog of X-ray sources \citep{pavlinsky2022} in the first year of the \srg/\art\ survey and the photometric distance to the object calculated from its redshift. The presented values of $\lx$ disregard the $k$-corrections and were not corrected for absorption on the line of sight.

\begin{figure}
  \centering
    \includegraphics[width=1\columnwidth]{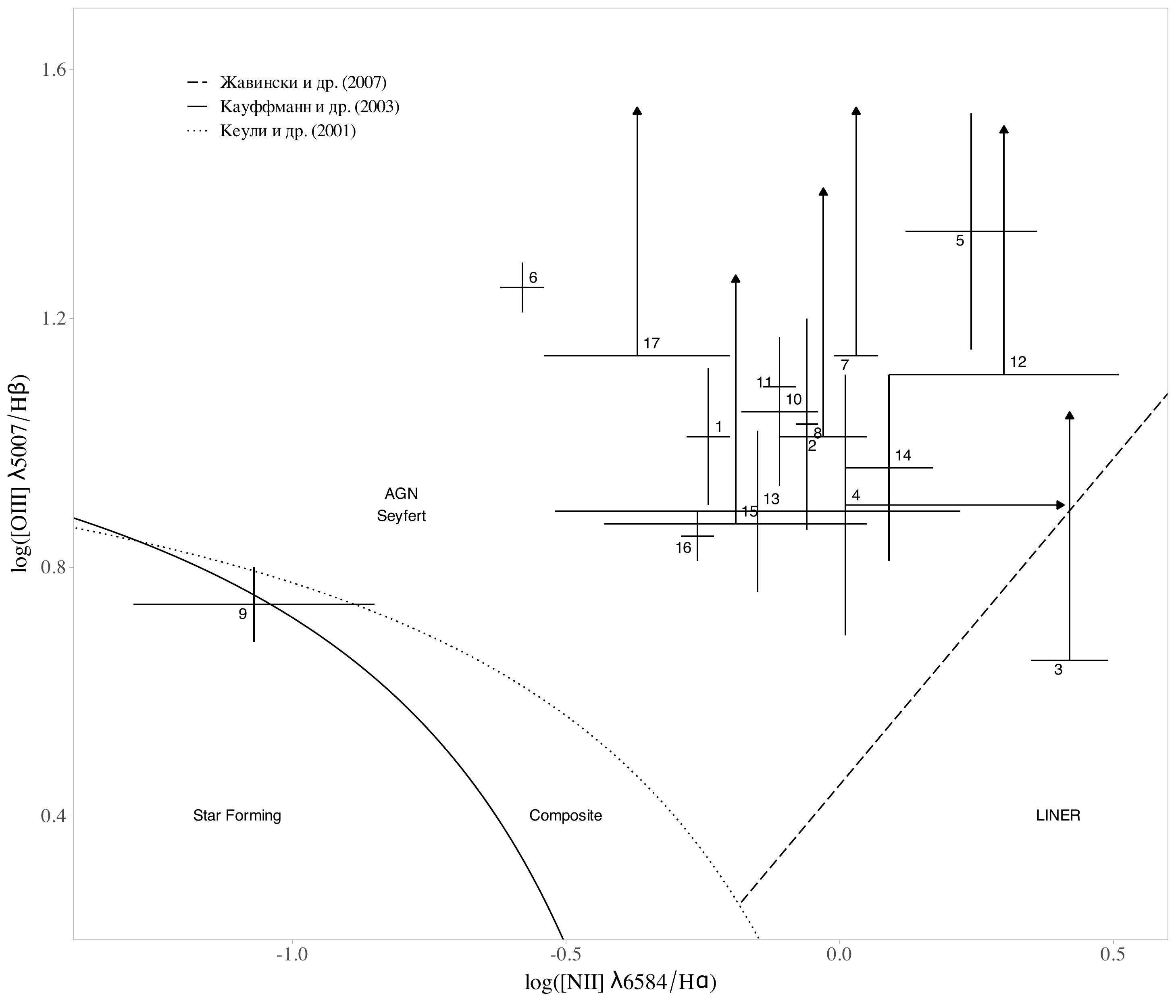}
    \caption{
    Locations of the AGNs being investigated on the BPT diagram \citep{baldwin1981} constructed from SDSS data (release 7, \citealt{abazajian2009}). 
    The confidence intervals of the flux ratios are presented on the diagram. 
    The arrow indicates the 2$\sigma$ lower limits. 
    The demarcation lines between different classes of galaxies were taken from \citealt{kauffmann2003} -- the solid line, 
    \citealt{kewley2001} -- the dotted line, 
    and \citealt{schawinski2007} -- the dashed line. 
    The sources are marked by the numbers from Tables~\ref{tab:list_src} and~\ref{tab:list_src_6df}.
  }
  \label{chart:bpt}
\end{figure}

All of the objects being discussed turned out to be nearby Seyfert galaxies with luminosities $\lx\sim 3\times 10^{43}$ -- $3\times 10^{44}$ erg/s and fall into the region of Seyfert galaxies on the standard BPT diagram (Fig.~\ref{chart:bpt}) of 
[OIII]$\lambda$5007/H${\beta}$ and [NII]$\lambda$6584/H${\alpha}$ flux ratios. 
Although the sources SRGA\,J$070636.4\!+\!635109$ and SRGA\,J$235250.6\!-\!170449$ are at the boundary of this region, the presence of broad Balmer line components in the spectra of these galaxies unambiguously suggests that these are Seyfert 1 galaxies.

\begin{figure}
  \centering
    \includegraphics[width=1\columnwidth]{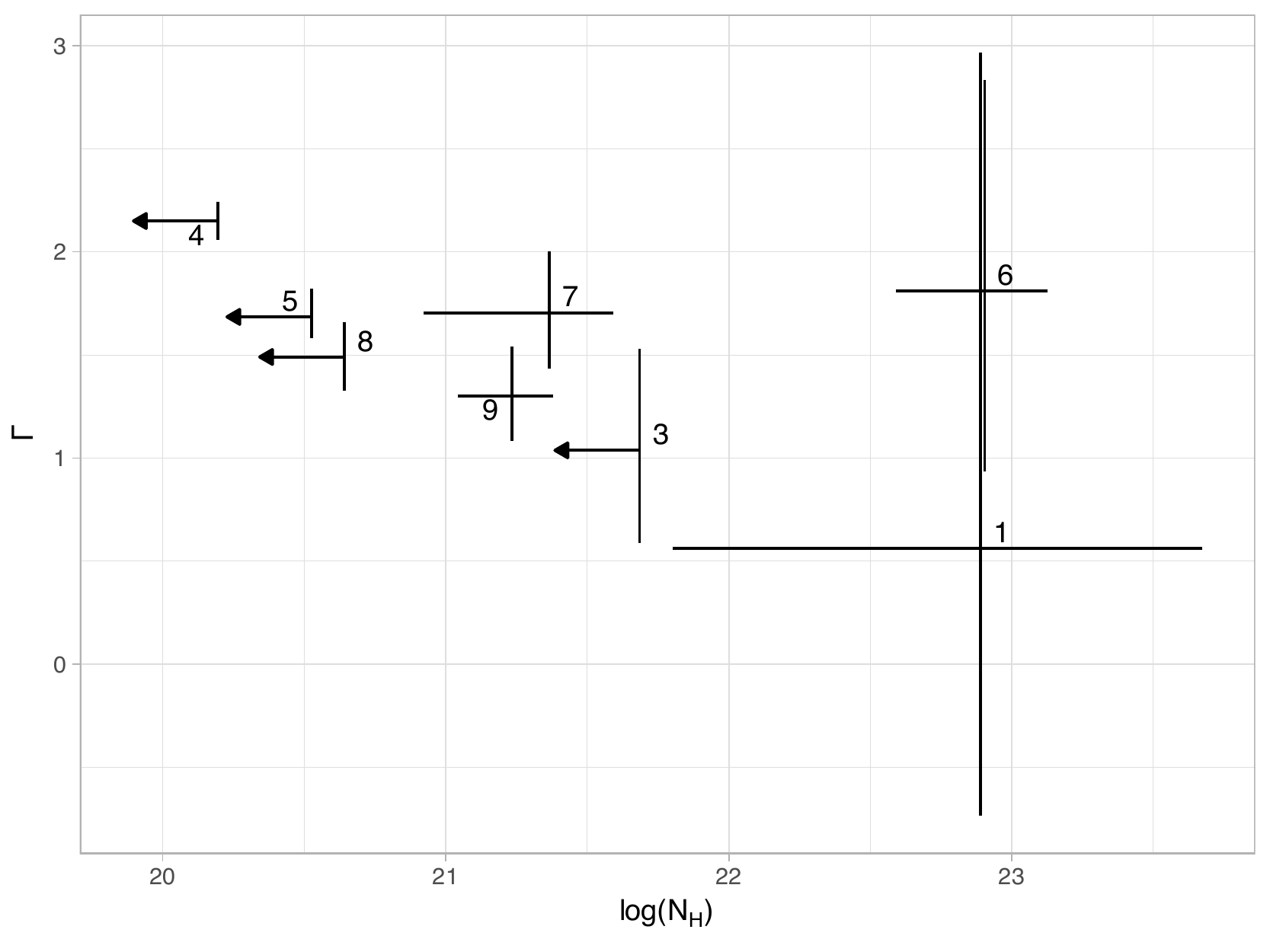}
    \caption{
    Slope of the X-ray power-law continuum versus intrinsic absorption column density for the eight AGNs investigated based on the \art\ and \ero\ data (see Table~\ref{tab:xray_params}). The arrows indicate the 90\% upper limits. The sources are indicated by the numbers from Table~\ref{tab:list_src}.
  }
  \label{chart:xray}
\end{figure}

In Fig.~\ref{chart:xray} the slope of the power-law continuum $\Gamma$ is plotted against the intrinsic absorption column density $\nh$ for the eight objects from our sample for which the X-ray spectra from the \art\ and \ero\ data were analyzed. Most of the derived slopes agree within the error limits with $\Gamma\sim 1.5$--2, which is typical for AGNs.

\begin{table*}
  \caption{
  Properties of the AGNs whose spectra were taken at the $\azt$ and $\rtt$ telescopes.
  } 
  \label{tab:list2}
  \vskip 2mm
  \renewcommand{\arraystretch}{1.1}
  \renewcommand{\tabcolsep}{0.35cm}
  \centering
  \footnotesize
  \begin{tabular}{cccc}
    \noalign{\doubleline}     
    Object & Optical type & $z$ &  $\log\lx$$^1$ \\
  \hline
  SRGA\,J$025234.3\!+\!431004$ & Sy2 & $0.05123\pm 0.00024$   & $43.0 \pm 0.4$ \\
  SRGA\,J$062627.2\!+\!072734$ & Sy1 & $0.04254\pm 0.00013$   & $43.0 \pm 0.4$ \\
  SRGA\,J$070636.4\!+\!635109$ & Sy1.8 & $0.01404\pm 0.00019$ & $42.4 \pm 0.3$ \\
  SRGA\,J$092021.6\!+\!860249$ & Sy1 & $0.05286\pm 0.00013$   & $43.5 \pm 0.2$ \\
  SRGA\,J$195702.4\!+\!615036$ & Sy1 & $0.05857\pm 0.00014$   & $43.4 \pm 0.2$ \\
  SRGA\,J$221913.2\!+\!362014$ & Sy2 & $0.14667\pm 0.00003$   & $44.4 \pm 0.2$ \\
  SRGA\,J$223714.9\!+\!402939$ & Sy1 & $0.05818\pm 0.00011$   & $43.6 \pm 0.3$ \\
  SRGA\,J$232037.8\!+\!482329$ & Sy2 & $0.04197\pm 0.00017$   & $42.8 \pm 0.4$ \\
  SRGA\,J$235250.6\!-\!170449$ & Sy1 & $0.05502\pm 0.00012$   & $43.7 \pm 0.3$ \\
    \noalign{\vskip 3pt\hrule\vskip 5pt}
  \end{tabular}
  \begin{flushleft}
  $^1$ The luminosity in the observed 4--12~keV energy band in units of erg/s uncorrected for absorption. The error corresponds to the 68\% confidence interval.
  \end{flushleft}
 \end{table*}

\begin{table*}
  \caption{
  Properties of the AGNs whose spectra were taken during the 6dF survey.
  } 
  \label{tab:list2_6df}
  \vskip 2mm
  \renewcommand{\arraystretch}{1.1}
  \renewcommand{\tabcolsep}{0.35cm}
  \centering
  \footnotesize
  \begin{tabular}{cccc}
    \noalign{\doubleline}     
    Object & Optical type & $z^1$ &  $\log\lx$$^2$ \\
  \hline
  SRGA\,J$030838.1\!-\!552041$ & Sy2   & 0.07791 & $43.8 \pm 0.2$ \\
  SRGA\,J$052959.8\!-\!340157$ & Sy1.8 & 0.07900 & $43.8 \pm 0.2$ \\
  SRGA\,J$055053.7\!-\!621457$ & Sy1   & 0.05875 & $43.0 \pm 0.2$ \\
  SRGA\,J$060241.1\!-\!595152$ & Sy2   & 0.10051 & $43.8 \pm 0.2$ \\
  SRGA\,J$061322.9\!-\!290027$ & Sy2   & 0.07051 & $44.1 \pm 0.1$ \\
  SRGA\,J$063324.9\!-\!561424$ & Sy2   & 0.04784 & $43.2 \pm 0.2$ \\
  SRGA\,J$064421.5\!-\!662620$ & Sy1   & 0.07843 & $42.9 \pm 0.4$ \\
  SRGA\,J$072823.5\!-\!440823$ & Sy1   & 0.08171 & $43.8 \pm 0.2$ \\
    \noalign{\vskip 3pt\hrule\vskip 5pt}
  \end{tabular}
  \begin{flushleft}
  $^1$ The values were taken from the catalog of redshifts of the \emph{6dF} survey.
  $^2$ The luminosity in the observed 4--12~keV energy band in units of erg/s uncorrected for absorption. The error corresponds to the 68\% confidence interval, without the error in $z$.
  \end{flushleft}
 \end{table*}

\section{CONCLUSIONS}

Using the observations carried out at the AZT-33IK and RTT-150 telescopes and the archival spectroscopic data from the 6dF survey, we managed to identify 17 new AGNs among the X-ray sources detected during the first year of the \srg/\art\ all-sky survey. All of them turned out to be nearby Seyfert galaxies (eight Sy1, two Sy1.8, and seven Sy2) at redshifts from $z=0.014$ to $z=0.147$.

For eight objects located on the half of the sky $0<l<180^\circ$, 
we constructed broadband (0.2--20~keV)
X-ray spectra based on data from the \art\ and
\ero\ telescopes onboard the \srg observatory. 
An intrinsic absorption was revealed in the
spectra of three of these objects, with two of them
(the Seyfert 2 galaxies SRGA J025234.3+431004
and SRGA J221913.2+362014) being strongly absorbed ($\nh\sim 10^{23}$ cm$^{-2}$) and the third one (Sy1 SRGA J235250.6-70449) being characterized by a
relatively weak absorption ($\nh\sim 10^{22}$ cm$^{-2}$). Note H
that the strongly absorbed source SRGA J025234.3+ 431004 is associated with an edge-on galaxy (LEDA~90641). Therefore, absorption can arise in this case not only in the gas–dust torus around the supermassive black hole, but also in the interstellar medium of the galaxy.

The \srg all-sky survey continues. 
It is expected that by the end of the four-year-long survey the \art\ telescope will detect ~ 5000 sources in the 4--12~keV energy band, mostly AGNs at low redshifts \citep{pavlinsky2022}, including many previously unknown ones. As was demonstrated in this paper and the previous paper from this series \citep{zaznobin2021}, the problem of identifying new AGNs from the \srg/\art\ survey can be efficiently solved with 1.5-m optical telescopes.

\section*{Acknowledgements}

This work was supported by RSF grant no. 19-12-00396. We thank T\"{U}BITAK, the Space Research Institute of the Russian Academy of Sciences, the Kazan Federal University, and the Academy of Sciences of Tatarstan for supporting the observations at the Russian–Turkish 1.5-m telescope (RTT-150). The measurements with the AZT-33IK telescope were supported by the Ministry of Education and Science of the Russian Federation and were obtained using the equipment of the Angara sharing center\footnote{http://ckp-rf.ru/ckp/3056/}. In this study we used observational data from the \art\ and \ero\ telescopes onboard the \srg observatory. The \srg\ observatory was built by Roskos- mos in the interests of the Russian Academy of Sciences represented by its Space Research Institute (IKI) within the framework of the Russian Federal Space Program, with the participation of the Deutsche Zentrum für Luft- und Raumfahrt (\textit{DLR}). The \srg\ spacecraft was designed, built, launched, and is operated by the Lavochkin Association and its subcontractors. The science data are downlinked via the Deep Space Network Antennae in Bear Lakes, Ussuriysk, and Baykonur, funded by Roskosmos. The \ero\ X-ray telescope was built by a consortium of German Institutes led by MPE, and supported by \textit{DLR}. The \ero\ data used in this work were processed using the \textit{eSASS} software developed by the German \ero\ consortium and the proprietary data reduction and analysis software developed by the Russian \ero\ Consortium.

\bibliographystyle{mnras}
\bibliography{references}

\end{document}